\documentclass[
  reprint,
  superscriptaddress,
  amsmath,amssymb,
  aps,
  prb,
  longbibliography
]{revtex4-2}
\usepackage{amsmath}
\usepackage{placeins}  
\usepackage{float}    
\usepackage{array}
\usepackage{physics}
\usepackage{microtype}
\usepackage{xcolor}
\usepackage{graphicx}
\usepackage{svg}
\usepackage{dcolumn}
\usepackage{bm}
\usepackage{siunitx}
\usepackage[utf8]{inputenc}
\usepackage{lmodern}   
\usepackage[table]{xcolor}
\usepackage{colortbl}
\usepackage[T1]{fontenc}
\usepackage{hyperref}
\usepackage{multirow}
\usepackage{textgreek}
\usepackage{booktabs}
\usepackage{threeparttable}
\definecolor{darkgreen}{rgb}{0.0,0.45,0.0}

\hypersetup{
  colorlinks=true,
  linkcolor=black,
  citecolor=blue,
  urlcolor=blue
}
\definecolor{dark green}{RGB}{0,110,0}

\begin{document}

\title{ Negative thermal expansion, lattice dynamics, and complex magnetism in TbFeO$_3$}

\author{Shubham Farswan}
\email{phz238329@physics.iitd.ac.in}
\affiliation{Department of Physics, Indian Institute of Technology Delhi, New Delhi 110016, India}

\author{Reshma Kumawat}
\affiliation{Department of Physics, Indian Institute of Technology Delhi, New Delhi 110016, India}

\author{Dipankar Sarkar}
\affiliation{School of Physical Sciences, Indian Association for the Cultivation of Science, 2A \& 2B Raja S. C. Mullick Road, Kolkata 700032, India}

\author{Deeksha Singh}
\affiliation{Department of Physics, Indian Institute of Technology Delhi, New Delhi 110016, India}

\author{Md.\ Atif Hasan}
\affiliation{Department of Physics, Indian Institute of Technology Delhi, New Delhi 110016, India}

\author{Devajyoti Mukherjee}
\affiliation{School of Physical Sciences, Indian Association for the Cultivation of Science, 2A \& 2B Raja S. C. Mullick Road, Kolkata 700032, India}

\author{Kaushik Sen}
\email{kaushik.sen@physics.iitd.ac.in}
\affiliation{Department of Physics, Indian Institute of Technology Delhi, New Delhi 110016, India}

\date{\today}

\begin{abstract}
We report a temperature-dependent investigation of orthoferrite TbFeO$_3$ using x-ray diffraction, DC magnetization, and Raman scattering, complemented by room-temperature x-ray photoelectron spectroscopy. X-ray diffraction reveals negative thermal expansion over 5-300 K, with a small but systematic increase in unit-cell volume upon cooling in the absence of any structural phase transition. Raman scattering measurements identify the Raman-active phonon modes and show clear deviations from the conventional Klemens anharmonic decay model, particularly in phonon frequencies, indicating the presence of spin-phonon coupling. Two modes of $A_g$ and $B_{1g}$ symmetry exhibit a crossover from Gaussian-dominated line shapes at low temperatures to mixed Gaussian-Lorentzian profiles at higher temperatures, reflecting a transition from inhomogeneous broadening to lifetime-driven dynamics. High-energy Raman spectra reveal two-magnon excitations associated with the Fe sublattice, consistent with linear spin-wave theory, whose spectral weight shows only weak temperature dependence. In addition, a broad Raman mode emerging below $\sim 175$ K exhibits an onset-like temperature evolution and coincides with the occurrence of phonon anomalies, while no corresponding strong anomaly is observed in the two-magnon response. Taken together, these results establish TbFeO$_3$ as a system with pronounced interplay among lattice dynamics, spin correlations, and emergent local magnetic-lattice anomalies.
\end{abstract}

\maketitle

\section{Introduction}
Complex oxides continue to attract broad interest because of the strong interplay among charge, spin, orbital, and lattice degrees of freedom \cite{Tokura2000,Imada1998}, which gives rise to a wide range of emergent phenomena such as colossal magnetoresistance \cite{DAGOTTO20011}, high-temperature superconductivity \cite{Keimer2015}, and multiferroicity \cite{Cheong2007,Ngai2014}. In particular, materials that host coupled magnetic and lattice responses provide an important setting for examining how microscopic interactions manifest in measurable structural and spectroscopic properties. Among such systems, rare-earth orthoferrites, RFeO$_3$ (R = rare-earth elements, Y, Sc), constitute an important family of correlated oxides with rich and tunable magnetic behavior \cite{White1969}.

The orthoferrites crystallize in the orthorhombic \textit{Pbnm} structure and contain two magnetic sublattices, Fe$^{3+}$ (3$d^5$) and R$^{3+}$ (4$f^n$) \cite{Gordon1976}. Their magnetism is governed by three principal exchange interactions: Fe$^{3+}$-Fe$^{3+}$, Fe$^{3+}$-R$^{3+}$, and R$^{3+}$-R$^{3+}$. A common feature of this family is canted antiferromagnetism driven by the antisymmetric Dzyaloshinskii-Moriya (DM) interaction, which generates a weak ferromagnetic moment along specific crystallographic directions \cite{White1969,Nikolov1994}. The Fe$^{3+}$ sublattice orders in a G-type antiferromagnetic configuration below the N\'eel temperature $T_{\mathrm{N}} \approx 623$-$740$~K, depending on the rare-earth ion, while the R$^{3+}$ moments order at much lower temperatures. The coexistence of these magnetic subsystems leads to spin-reorientation phenomena and pronounced magnetostructural coupling \cite{WeberNature2022}. Recent predictions of topological magnon excitations in RFeO$_3$ have further renewed interest in orthoferrites as model systems for studying collective magnetic excitations in correlated oxides \cite{Karaki2023,Karaki2022_arxiv}.

In addition to their magnetic richness, several orthoferrites exhibit signatures of magnetically induced ferroelectricity, where electric polarization emerges as a consequence of specific spin configurations rather than a conventional structural instability \cite{Tokunaga2008,Tokunaga2009,Lee2011}. Although the orthorhombic \textit{Pbnm} structure is centrosymmetric and therefore incompatible with proper ferroelectricity, inversion symmetry can be broken by magnetic ordering at low temperatures, giving rise to type-II multiferroic behavior. Such effects have been reported, for example, in GdFeO$_3$ \cite{Tokunaga2009} and DyFeO$_3$ \cite{Tokunaga2008}. In SmFeO$_3$ the polarization arises due to canted antiferromagnetic ordering of two nonequivalent spin pairs \cite{Lee2011}.  In TbFeO$_3$, magnetic-field-induced electric polarization and memory effects have also been reported~\cite{Ivanov2023,Ivanov2023_2}. In addition, a dielectric anomaly has been reported near 200 K, and ferroelectric ordering near 200 K and 2 K has recently been observed in single crystals~\cite{Indra2025}, further motivating a detailed study of its coupled lattice and magnetic properties.

Despite this broad interest, the relationships among thermal expansion, lattice dynamics, magnetic correlations, and local inhomogeneity in TbFeO$_3$ remain incompletely understood. Previous Raman studies have established the phonon spectrum and symmetry assignments and have reported temperature-dependent phonon anomalies in TbFeO$_3$ and related orthoferrites \cite{Vilarinho2022,Dubrovin2024}. However, these spectroscopic observations have not been systematically examined together with the temperature evolution of the crystal lattice and magnetic response over a broad temperature range. Consequently, it remains unclear whether the reported phonon anomalies are isolated mode-specific effects or are part of a broader spin-lattice response. Raman scattering is particularly useful for addressing this question because it simultaneously probes first-order phonons, higher-order phonon processes, and magnetic excitations such as two-magnon scattering, thereby providing a direct window into spin-lattice coupling \cite{Sen2019,Sen2020,Devereaux2007}.

In this work, we investigate polycrystalline TbFeO$_3$ by combining temperature-dependent x-ray diffraction, DC magnetization, and Raman scattering, complemented by room-temperature x-ray photoelectron spectroscopy. This combined approach enables us to examine the structural evolution, magnetic response, and lattice dynamics within a common experimental framework. We establish a small negative thermal expansion in the absence of a structural phase transition and identify mixed-valence states at the sample surface. Several Raman-active phonon modes deviate from the conventional Klemens anharmonic decay model, indicating contributions beyond simple phonon-phonon scattering. Two phonon modes exhibit a crossover from Gaussian-dominated to mixed Gaussian-Lorentzian line shapes, reflecting a temperature-dependent change in the phonon-broadening mechanism. At higher energies, we identify two-magnon Raman excitations associated with the Fe sublattice, supported by linear spin-wave calculations. In addition, a broad Raman feature emerges below $\sim 175$~K and shows a gradual temperature evolution. Taken together, these results establish the relevant temperature scales and reveal correlated structural, vibrational, and magnetic responses in TbFeO$_3$, providing a unified experimental basis for discussing spin-lattice coupling and local magnetic inhomogeneity.


\section{Experimental details}

\subsection{Solid-state Synthesis}
    We grew polycrystalline TbFeO$_3$ using solid-state synthesis. High-purity precursor oxides, Tb$_4$O$_7$ (99.9\%, Thermo Fisher) and Fe$_2$O$_3$ (99.9\%, Thermo Fisher), were weighed in their stoichiometric proportions and homogenized by grinding for approximately 3 hours. The homogenized mixture was subsequently heated at 1100 °C for 12 h in furnace, employing a heating rate of 3 °C min$^{-1}$. The resulting powder was reground and mixed twice to eliminate any residual precursor phases. The powder was then pressed into pellets for Raman and X-ray diffraction (XRD) measurements.

\subsection{Structural and Compositonal Characterization}
    We performed symmetric $\theta$-2$\theta$ X-ray diffraction (XRD) measurements using a Rigaku SmartLab 9 kW XG diffractometer equipped with a five-axis goniometer in a collimated parallel-beam configuration. The measurements were carried out by employing Cu K$\alpha_1$ radiation with a wavelength of $\lambda = 1.5406\,\text{\AA}$. Temperature-dependent XRD measurements were conducted in the range of 5-300~K. Rietveld refinement of the diffraction patterns was performed using the FullProf suite~\cite{RodriguezCarvajal1990} to extract the lattice parameters and structural information at every measured temperature. The XRD data was recorded in the range $10^{\circ}$ $\leq$  $\theta$ $\leq$ $90^{\circ}$. {\par}
    The room temperature x-ray photoelectron spectroscopy (XPS) measurements were performed using the Kratos Analytical Ltd (AXIS SUPRA model) spectrometer with monochromatic Al K$\alpha$ (h$\nu$=1486.6 eV) source and energy resolution of 0.5\,eV. All the core level spectra are calibrated using the Carbon $1s$ peak at 284.8\,eV as a reference. For the analysis of XPS, inelastic background subtraction was performed and the core-levels then fitted using the Gaussain Lorentzain product lineshape (GL(60)) as defined in CASAXPS \cite{Fairley2009} which include contribution from Gaussian and Lorentzian profiles. 

\subsection{Magnetization measurements}
    We measured DC magnetization ($M$) of polycrystalline TFO as a function of temperature ($T$) and magnetic field ($\mu_0H$) using a Quantum Design made MPMS-3 SQUID magnetometer. For measurements, we wrapped $13$\,mg of TFO powder in teflon tape, and mounted inside a plastic straw. Subsequently, the sample was cooled to $2$\,K from room temperature at $\mu_0H=500$\,Oe. Field-cooled (FC) magnetization data were recorded  while warming up from $2$ to $380$\,K. 
    For high-temperature magnetization measurements, an additional 13\,mg TFO powder was mixed with zircon cement. The sample was initially heated to 750~K in absence of magnetic field. Afterwards, the sample was cooled to $300$\,K under the same magnetic field of  $\mu_0H=500$\,Oe. Subsequently, FC magnetization data were recorded while warming from $300$ to $750$\,K. 

\subsection{Raman Scattering}
    Temperature-dependent raman spectroscopic measurements were carried out using a home-built setup consisting of a Horiba spectrometer (iHR550) integrated with a closed-cycle helium cryostat. A He-Ne laser having wavelength of 632.8\,nm was used for excitation, and backscattering geometry is used to collect the Raman signal. The sample was mounted on a copper holder using Apiezon grease, while silver paste was applied to ensure good thermal contact between the sample surface and the copper block attached to cryostat cold finger. The scattered light was focused using a long working-distance objective lens (10$\times$ Mitutoyo, $f=200$\,mm) on to the sample. The elastic part of spectrum was suppressed using two notch filters. The scattered light was dispersed by a 1200 grooves/mm grating, providing a spectral resolution of approximately 2.5 \,cm$^{-1}$ and detected using a CCD detector.

    \begin{figure*}[t]
        \centering
        \includegraphics[width=1\linewidth]{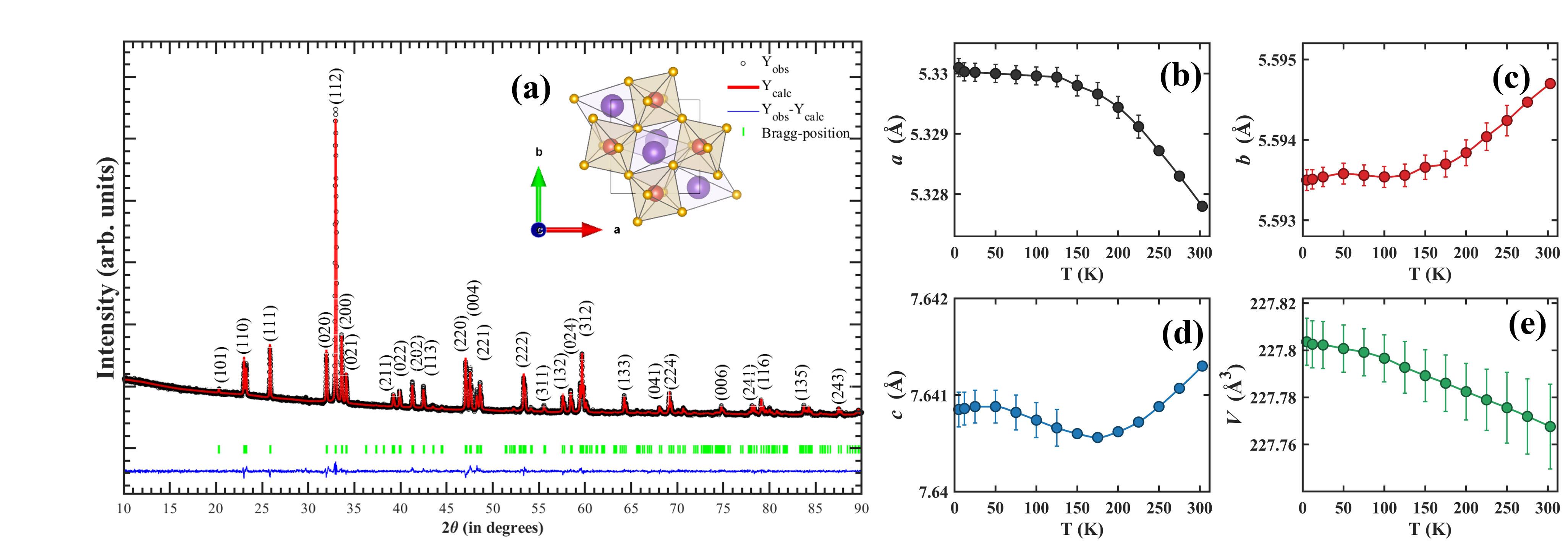}
        \caption{Structural refinement of TbFeO$_3$ x-ray diffraction pattern at room temperature. Unit cell, view along the c-axis. Here Tb atoms are shown in purple spheres, Fe atoms as red spheres, and O atoms as yellow spheres (b-d) Temperature evolution of the refined lattice parameters $a(T)$, $b(T)$, and $c(T)$ respectively extracted by Rietveld refinement of temperature-dependent XRD measurements. (e) Temperature dependence of unit-cell volume $V(T)$, indicating negative thermal expansion.The error bars represent the standard uncertainties obtained from the Rietveld refinement.}
        \label{fig:xrd}
    \end{figure*}

    For temperature-dependent measurements, the sample was cooled down to 11\,K, and Raman spectra were recorded in the spectral range of 38-840\,cm$^{-1}$ at several temperatures during warming up to 300\,K. A minimal laser power of less than 1\,mW was used to avoid laser-induced local heating.
    The measured Raman scattering intensity is directly related to the imaginary part of the dynamical susceptibility, $\chi''(\omega)$, through the expression
    \begin{equation}
        I(\omega) \propto \chi''(\omega)\,[n(\omega)+1],
        \label{eq:StokesIntensity}
    \end{equation}
    where $n(\omega)$ denotes the phonon occupation number described by the Bose-Einstein distribution,
    \[
        n(\omega)=\left(e^{\hbar\omega/k_{\mathrm{B}}T}-1\right)^{-1}.
    \]
    To enable a consistent comparison of spectra obtained at different temperatures, the measured Raman intensities were normalized by the Bose population factor $[n(\omega)+1]$, ensuring that the resulting spectra are directly proportional to $\chi''(\omega)$. The resulting Bose-corrected spectra were analyzed by fitting the individual phonon modes using Voigt line shapes, where the Gaussian component accounts for the instrumental broadening and the Lorentzian component represents the intrinsic phonon lifetime. An instrumental resolution of $2.5\,\mathrm{cm^{-1}}$ was considered in the fitting procedure. The background contribution was modeled as a combination of a quadratic term, a quasi-elastic peak, and a logistic continuum. The complete Raman response at each temperature was obtained through simultaneous fitting of both the phonon modes and the background components. 
    

\section{Results and discussion}

\subsection{Structural Characterization}
    Fig.~\ref{fig:xrd}(a) shows the powder x-ray diffraction (XRD) pattern of polycrystalline TFO measured at room-temperature. All Bragg peaks match the reported XRD pattern of TFO for JCPDS PDF No.~47-0068 \cite{Bombik2003}, which confirms that the compound crystallizes in an orthorhombic structure with the space group of \textit{Pbnm}. Table~\ref{tab:lattice_parameters} lists the lattice parameters obtained from the Rietveld refinement ($\chi^{2}=1.44$) of the XRD pattern using the space group of \textit{Pbnm}.  These values are in good agreement with previously reported results, as given in Table~\ref{tab:lattice_parameters}.
    
\begin{table}[htbp]  
    \caption{\label{tab:lattice_parameters}
    Refined room-temperature lattice parameters of TbFeO$_3$
    compared with reported values. All values are in \AA.}
    \centering
    \setlength{\tabcolsep}{4pt}
    \renewcommand{\arraystretch}{1.1}
    \begin{ruledtabular}
        \begin{tabular}{lcccc}
            Parameter
            & This work
            & Ref. \cite{Dubrovin2024}
            & Ref. \cite{Marezio1970} 
            & Ref. \cite{Ovsianikov2022}\\
            \hline
            $a$ & 5.32  & 5.33 & 5.326 & 5.3273(2)\\
            $b$ & 5.59  & 5.60 & 5.602 & 5.5996(3) \\
            $c$ & 7.64  & 7.65 & 7.635 & 7.6404(4)\\
        \end{tabular}
    \end{ruledtabular}
\end{table}

     \begin{table*}[t]
        \caption{Representative materials exhibiting negative thermal expansion (NTE) and their microscopic mechanisms.}
        \begin{ruledtabular}
            \begin{tabular}{ccccc}
                Compound & Origin of NTE & $T$ range (K) & Volume change (-$\Delta V/V$,\%) & Ref. \\
                \hline
                Bi(Ni,Fe)O$_3$ & Charge transfer & 320-370 & $\sim 2$ & \cite{Nishikubo2019} \\
                Bi$_{1-x}$La$_x$NiO$_3$ & Charge transfer & 320-380 & $\sim 2$ & \cite{Azuma2011} \\
                Mn(Co,Cr)Ge & Magnetic ordering & 122-332 & $\sim 3.2$ & \cite{Zhao2015} \\
                La(Fe,Co,Si)$_{13}$ & Magnetic ordering & 240-350 & $\sim 1.1$ & \cite{Huang2013} \\
                LaMnO$_3$ & Orbital ordering (Jahn-Teller) & 700-750 & $\sim 0.36$ & \cite{Wdowik2011} \\
                PrMnO$_3$ & Orbital ordering (Jahn-Teller) & 900-1100 & $\sim 1.7$ & \cite{Qin2025} \\
                Ca$_2$(Ru,Cr)O$_4$ & Orbital ordering & 135-345 & $\sim 1$ & \cite{Qi2010} \\
                Cu$_2$V$_2$O$_7$ & Low-frequency phonons (RUM) & 98-475 & $\sim 0.38$ & \cite{Shi2020} \\
                BaTiO$_3$ & Ferroelectric transition & 393-403 & -- & \cite{Adhikary2023} \\
                Ca$_2$RuO$_4$ & Oxygen vacancy controlled & 145-345 & $\sim 6.7$ & \cite{Takenaka2017} \\
            \end{tabular}
        \end{ruledtabular}
        \label{tab:NTE_comparison}
    \end{table*}

Figs.~\ref{fig:xrd}(b–e) show the temperature dependence of the lattice parameters and the unit-cell volume in the range 5–303~K. In typical crystalline solids without structural phase transitions, lattice anharmonicity leads to a monotonic decrease of the unit-cell volume upon cooling, eventually approaching a nearly temperature-independent value at low temperatures (typically below $\sim 100$~K), consistent with a positive thermal expansion coefficient. For comparison, the relative volume contraction on cooling is material dependent: representative values are approximately 0.14\% for Al, 0.126\% for Cu~\cite{touloukian1975thermal}, and 0.006\% for Al$_2$O$_3$~\cite{lucht2003precise}.

In contrast to this conventional behavior, we observe an anomalous expansion of the unit-cell volume with decreasing temperature, as shown in Fig.~\ref{fig:xrd}(e). The rate of expansion decreases significantly below $\sim 100$ K, and the overall magnitude of the effect remains small, corresponding to a total volume increase of $\sim 0.017\%$ between 303 K and 5 K. 

A detailed analysis reveals an anisotropic and non-monotonic lattice evolution: the $a$-axis expands by $\sim 0.04\%$, while the $b$- and $c$-axes contract by $\sim 0.02\%$ and $\sim 0.005\%$ with decreasing temperature, respectively. Below $\sim 150$ K, the temperature dependence of all lattice parameters becomes significantly weaker. Despite this anomalous behavior, temperature-dependent XRD measurements show no evidence of a structural phase transition over the entire temperature range. {\par}

Negative thermal expansion (NTE) has been reported in a variety of transition-metal oxides, where it is often associated with electronically or magnetically driven lattice responses rather than pure octahedral rearrangement (see Table \ref{tab:NTE_comparison}). In different materials, several mechanisms have been discussed in the literature, including charge-transfer induced volume changes, magnetoelastic coupling via exchange striction, and orbital-ordering driven lattice distortions. For example, in manganites such as PrMnO$_3$ \cite{Qin2025} and related compounds, NTE has been linked to cooperative Jahn-Teller distortions and orbital ordering of Mn$^{3+}$ ions, which modify the MnO$_6$ octahedral network.
Recent studies have further highlighted the important role of oxygen stoichiometry in governing NTE. In PrMnO$_3$, stoichiometric and oxygen-rich samples exhibit markedly different thermal expansion behaviour, demonstrating that oxygen stoichiometry can significantly modify the local structure and electronic configuration responsible for NTE \cite{Qin2025}. Likewise, in reduced Ca$_2$RuO$_{4-\delta}$, oxygen non-stoichiometry has been shown to tune the magnitude of the NTE through changes in the orbital-ordered state \cite{Takenaka2017}. These examples illustrate that oxygen non-stoichiometry can influence the microscopic mechanisms underlying NTE in transition-metal oxides.
Magnetically driven NTE has also been reported in systems where strong exchange interactions lead to volume changes through spin-lattice coupling \cite{Zhao2015,Huang2013}.

For TbFeO$_3$, which adopts the orthorhombic $Pbnm$ structure, negative thermal expansion is established experimentally over a wide temperature range. The present work does not attempt to identify a unique microscopic origin; instead, we emphasize that multiple mechanisms discussed in related systems, including magnetoelastic effects and electronically driven bonding changes, could be operative. The presence of negative thermal expansion further motivates a detailed study of the lattice dynamics in TFO, for which Raman scattering provides a direct microscopic probe. {\par}  
\subsection{XPS Core-Level Spectra and Chemical Environment}
\begin{figure}[htbp]
        \centering
        \includegraphics[width=1\linewidth]{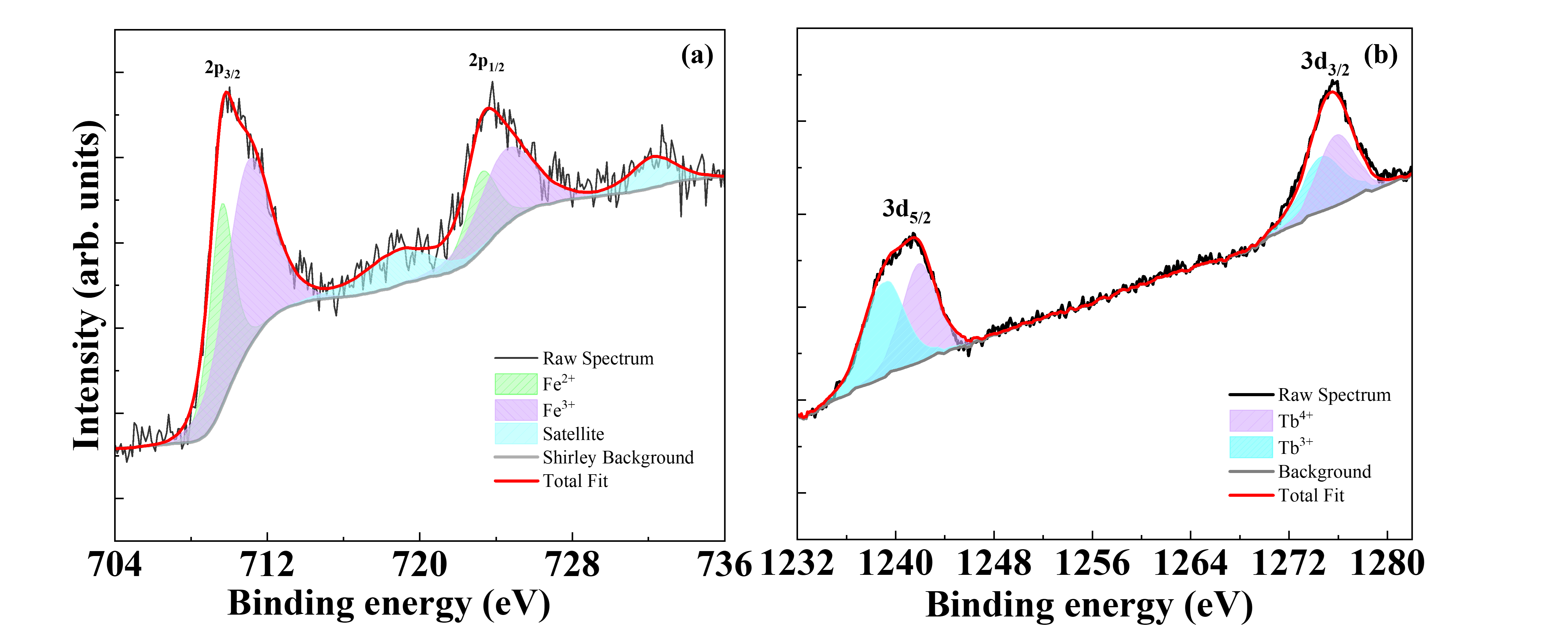}
        \caption{High-resolution XPS spectra of the Fe-$2p$ and Tb-$3d$ core levels of TbFeO$_3$, showing the characteristic spin-orbit split components along with deconvoluted contributions corresponding to Fe$^{2+}$, Fe$^{3+}$ and Tb$^{3+}$, Tb$^{4+}$ ionic states.}
        \label{fig:xps}
    \end{figure}

        X-ray photoemission spectroscopy (XPS)  was used to identify the oxidation states of Fe and Tb in TFO. Room-temperature XPS survey spectrum of TFO confirms the presence of all constituent elements (see Fig.~S4 in the SI).  The corresponding core-level spectra of Tb-3d and Fe-2p are presented in Figs.~\ref{fig:xps}(a) and (b), respectively. 
        Fe-2p spectrum shown in Fig.~\ref{fig:xps}(a) reveals two strong peaks belonging to the spin-orbit coupling induced doublets: Fe-2p$_{1/2}$ and Fe-2p$_{3/2}$. The characteristic Fe-2p$_{3/2}$ peak splits into two components located at 709.60 eV and 710 .98 eV and so is the Fe-2p$_{1/2}$ peak at 723.24 eV and 724.48 eV \cite{Bagus2021}.
        
     These split peaks can be assigned to the $+2$ and $+3$ oxidation states of Fe present at the surface of the sample. Peaks present at 714.55 eV and 732.21 eV are satellite peaks \cite{Yamashita2008,Bagus2021}. Similarly, Tb-3d spectrum shown in Fig.~\ref{fig:xps}(b) displays two spin-orbit doublets: Tb-3d$_{3/2}$ and Tb-3d$_{5/2}$. Each of these peaks further splits. In particular, Tb-3d$_{3/2}$ peak splits into two components located at 1274.87\,eV and 1275.81\,eV; and so is the Tb-3d$_{5/2}$ peak at 1238.87\,eV and 1241.77\,eV \cite{Zhu2016}.

        
        \begin{table*}
        \caption{Temperature-dependent evolution of magnetic ordering.}
        \label{tab:mag_ord}
        \begin{tabular}{cccccc}
        \toprule
        \textbf{Temperature Range} & T > 650K & T < 650K & T $\sim$ 280K & 4K < T < 8K \cite{Artyukhin2012} & T < 4K \cite{Artyukhin2012}\\ \hline
        &   &  Fe spins gets   & Magnetization  & Fe SR from & Fe AF aligned\\
        \textbf{Magnetic Order}& Paramagnetic &  AF aligned with & Reversal & $G_xF_ z$ to $G_zF_x$ structure &  $G_xF_z$ structure and\\
        &  & weak canting & & and Tb SR from & Tb AF aligned\\
        &&&& $A_xG_y$ to $F_xC_y$ & $A_xG_y$ structure\\
        \hline\hline
        \end{tabular}
    \end{table*}
        
        Thus, the XPS results confirm mixed Tb$^{3+}$/Tb$^{4+}$ and Fe$^{3+}$/Fe$^{2+}$ states at the sample surface. In stoichiometric TbFeO$_3$, the oxidation states of Tb, Fe, and O are $+3$, $+3$, and $-2$, respectively. The observed surface mixed valence indicates local non-stoichiometry, possibly associated with oxygen vacancies. However, because XPS is surface sensitive, these results do not establish the oxygen deficiency or valence fractions in the bulk. The surface fractions of Fe$^{2+}$ and Fe$^{3+}$ are 35.61\% and 64.38\%, respectively, while those of Tb$^{3+}$ and Tb$^{4+}$ are 49.51\% and 50.48\%, respectively. Details of the atomic-concentration analysis are provided in the SI.


\subsection{Magnetic Properties}
\begin{figure}[htbp]
        \centering
        \includegraphics[width=1\linewidth]{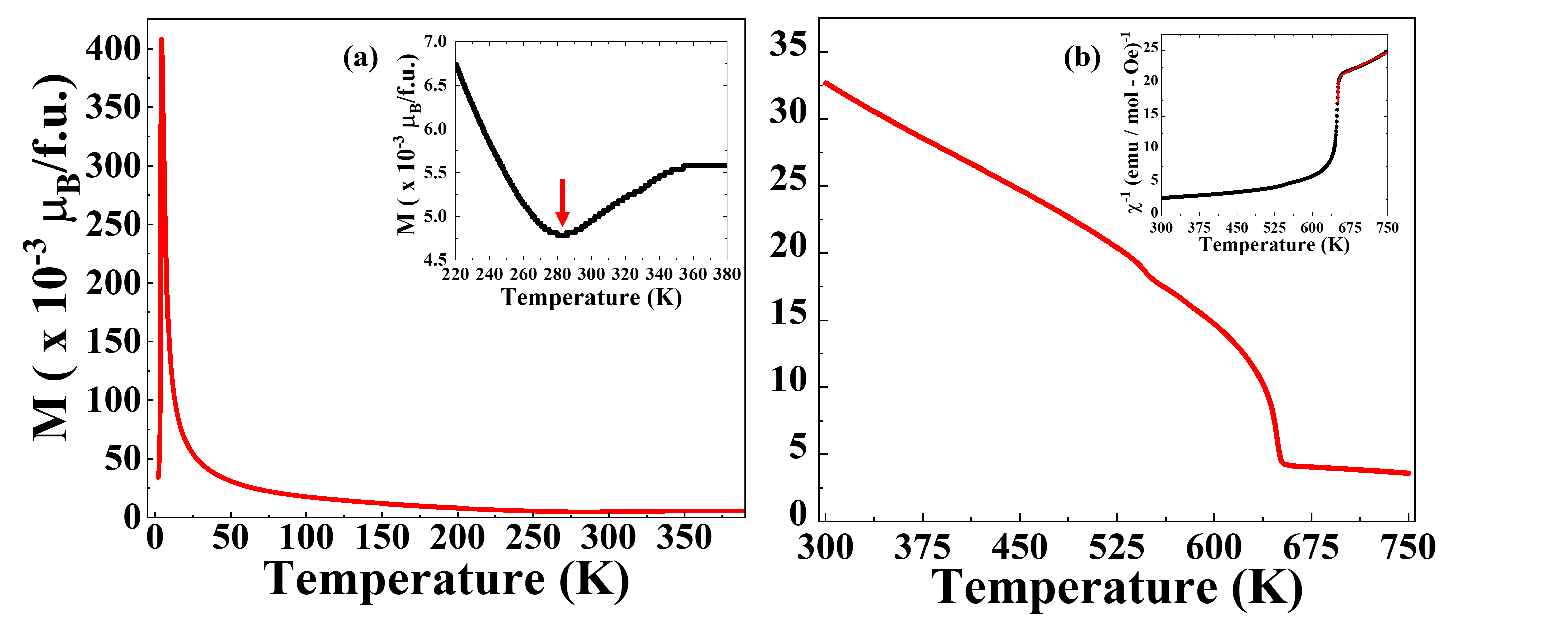}
        \caption{ Magnetization as a function of temperature for TbFeO$_3$ measured under an applied magnetic field of 500 Oe: (a) low-temperature regime and (b) high-temperature regime. The inset of panel (a) highlights the variation in magnetization in the vicinity of 280 K. The inset of panel (b) shows the inverse magnetic susceptibility fitted using the modified Curie–Weiss expression given in Eq. (\ref{eq:modified_curie_weiss}).}
        \label{fig:mt}
    \end{figure}

    Figs.~\ref{fig:mt} (a) and (b) show field-cooled $M$-$T$ data of polycrystalline TFO in low- and high-temperature regimes, respectively. In the high-temperature regime (Fig.~\ref{fig:mt}(b)), $M$ distinctly increases below $\sim650$\,K. However,
    the magnitude of $M$ remains very small, only of the order of $10^{-3}\mu_B$ per formula unit (f.u.) in contrast to the expected effective magnetic moment of 11.38 $\mu_B$/f.u. for stoichiometric TFO (We note that this comparison is only intended to illustrate the large difference between the measured magnetization and the ionic paramagnetic moment expected for stoichiometric TbFeO$_3$.). The magnetization remains weak ($10^{-1}\mu_B$/f.u.) even under high magnetic field of 5 T down to 200~K (Fig.~S5 in the SI). This weak positive magnetization is consistent with the reported canted antiferromagnetic (AF) order of the Fe sublattice \cite{Kim2007}. In particular, Fe moments adopt a $G$-type AF configuration, with spin canting induced by the Dzyaloshinskii-Moriya (DM) interaction between Fe$^{3+}$ spins, resulting in a weak ferromagnetism \cite{Nikolov1994}. {\par}

    The presence of antisymmetric Dzyaloshinskii-Moriya (DM) interactions leads to corrections to the conventional Curie-Weiss behavior of ordered magnets, 
    \begin{equation}
        \chi = \frac{C}{T - \theta_{\mathrm{CW}}}.
    \end{equation}
    where $C$ is a constant and $\theta_{CW}$ is the Curie-Weiss temperature. Within Moriya’s treatment of weak ferromagnets, building on symmetry arguments for antisymmetric exchange, the magnetic susceptibility $(\chi)$ can be expressed in the following modified Curie-Weiss form \cite{Moriya1960,Dzyaloshinskii1958}:
    \begin{equation}
    \chi = \frac{C}{T - \theta_{\mathrm{CW}}} \cdot \frac{T - T_0}{T - T_N},
\label{eq:modified_curie_weiss}
\end{equation}
where $T_0$ is an additional parameter. The factor $(T-T_0)/(T-T_N)$ arises in the mean-field description of a canted antiferromagnet~\cite{Moriya1960}. Within this framework,
\begin{equation}
    \frac{T_N-T_0}{T_N} \simeq
    \frac{1}{2}\left(\frac{D}{J}\right)^2,
\label{eq:splitting}
\end{equation}
where $J$ and $D$ are the symmetric-exchange and DM coupling constants, respectively. Since $D/J \ll 1$ for a weakly canted antiferromagnet, the separation between $T_N$ and $T_0$ is expected to be very small. Using the reported canting parameters for TbFeO$_3$~\cite{Treves1965}, Eq.~\eqref{eq:splitting} gives $T_N-T_0 \approx 0.079$ K, which is below the resolution achievable from the present susceptibility data. The fit yields $T_0 = 651.31 \pm 0.14$ K and $T_N = 651.36 \pm 0.13$ K, corresponding to $T_N-T_0 = 0.05 \pm 0.19$ K. Thus, the splitting is unresolved within uncertainty and the factor $(T-T_0)/(T-T_N)$ provides little independent constraint on $T_0$. We therefore use Eq.~\eqref{eq:modified_curie_weiss} only as a phenomenological description of the susceptibility near $T_N$ and do not assign independent physical significance to the fitted value of $T_0$.
     
    From the fitted Curie constant $C = \dfrac{N_A \mu_{\mathrm{eff}}^2}{3k_B}$, we obtained the effective magnetic moment $\mu_{\mathrm{eff}} = 14.94\,\mu_B$ per f.u. The obtained $\theta_{\mathrm{CW}}$ is $59.6 \pm 5.4\,\mathrm{K}$. The positive sign of $\theta_{\mathrm{CW}}$ indicates the presence of ferromagnetic (FM) exchange correlations. Such FM correlations are embedded within the globally antiferromagnetic (AF) phase, characterized by a high N\'eel temperature of $T_N \sim 650$\,K. The origin of these FM correlations may be linked to the presence of multivalent magnetic ions, namely Fe$^{2+}$/Fe$^{3+}$ and Tb$^{3+}$/Tb$^{4+}$, which facilitate double exchange mechanisms \cite{zener1951interaction,Gennes1960}. {\par}
    We note that $\mu_{\mathrm{eff}}=14.94~\mu_B$/f.u. should be regarded as an apparent effective magnetic moment rather than the intrinsic ionic paramagnetic moment. Although oxygen non-stoichiometry can modify the magnetic response through mixed-valence $Tb^{3+}/Tb^{4+}$ and $Fe^{3+}/Fe^{2+}$ ions, the valence fractions obtained from XPS cannot be directly compared with the bulk magnetic susceptibility because XPS is a surface-sensitive probe. Furthermore, the Curie--Weiss fitting was performed over a limited temperature interval (650--750 K), close to the Fe magnetic ordering temperature, where short-range magnetic correlations, critical fluctuations, and exchange inhomogeneity are expected to cause deviations from ideal Curie-Weiss behaviour.

    In the low-temperature regime, as shown in Fig.~\ref{fig:mt} (a), the magnetization $\textbf{M}$ increases with decreasing temperature and exhibits a weak cusp (dip) around $\sim 280$~K, followed by a further increase upon cooling. Below $\sim 4$~K, $\textbf{M}$ shows a sharp decrease. Such a non-monotonic temperature dependence of $\textbf{M}$ in TFO is the result of two distinct magnetic sublattices, namely Fe and Tb \cite{Artyukhin2012}. In the high-temperature regime, below $T_N  \approx 650$~K, the magnetization is dominated by the Fe-sublattice. In this regime, dipolar interactions and single-ion anisotropy align the Fe spins in a \textbf{G}-type AF structure, with the AF propagation vector along $a$-axis and a weak FM magnetization component along $c$-axis \cite{Gordon1976}. Upon further cooling, the ordered Fe-sublattice generates an internal molecular field that progressively polarizes the Tb moments \cite{Yamaguchi1974}. Between $4~\mathrm{K}<T<8.5~\mathrm{K}$, the Fe sublattice adopts a \textbf{G}-type AF structure with the AF propagation vector along the $c$-axis and a weak FM component along the $a$-axis, while the Tb sublattice adopts a \textbf{C}-type AF structure with the AF propagation vector along the $b$-axis and a ferromagnetic component along the $a$-axis\cite{Artyukhin2012}. Below $\sim 4$~K, the Tb moments undergo a long-range AF ordering of \textbf{A}-type along a-axis and \textbf{G}-type along b-axis, accompanied by a spin-reorientation transition in the Fe sublattice back to high temperature order  \cite{Artyukhin2012} . 
    For brevity, Table \ref{tab:mag_ord} summarizes the overall magnetization behavior in different temperature ranges. Furthermore, the smooth and intrinsic nature of the M(T) response, without extraneous features does not reveal obvious signatures of magnetic impurity phases and supports the conclusion that the observed behavior arises from the inherent magnetic structure of TFO.
      \begin{figure*}[t]
        \centering
        \includegraphics[width=0.8\linewidth, keepaspectratio=true]{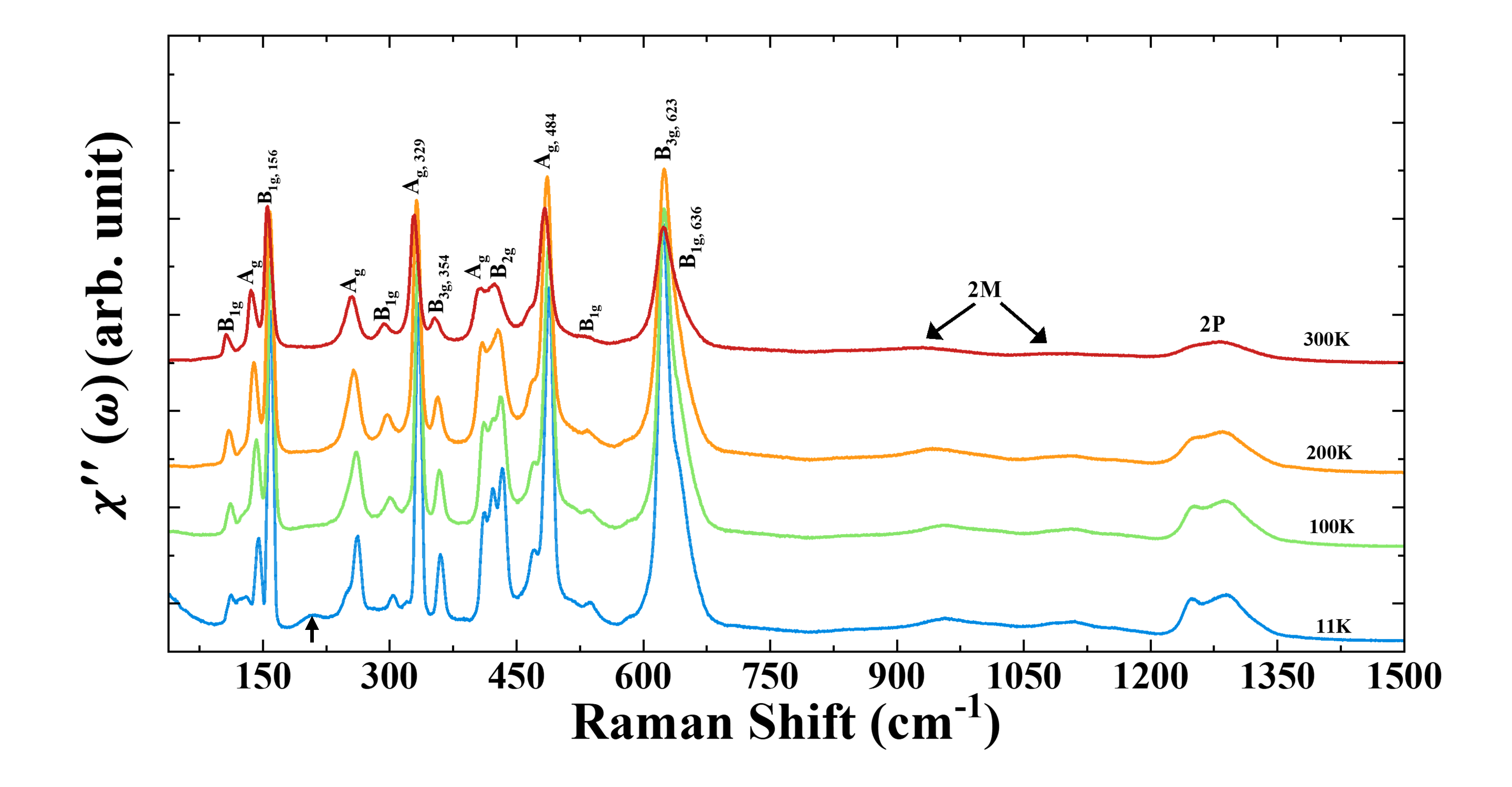}
        \caption{Temperature-dependent Raman spectra of polycrystalline TbFeO$_3$ measured at 11, 100, 200, and 300~K. Selected phonon modes are assigned, along with two-phonon (2P) and two-magnon (2M) excitations. The arrow indicates the emergence of a new mode at low temperatures.}
        \label{fig:tr1}
    \end{figure*}   
    
\subsection{Assignment of Raman-active modes arising from lattice dynamics and magnetism}
\textit{Phonon mode assignment.}
    Raman scattering is a photon-in, photon-out process that couples directly to electronic, spin, and lattice degrees of freedom, as well as to the interactions among them, including electron-phonon and spin-phonon coupling. In the insulating and magnetically ordered TFO, the Raman response $\left(\chi''(\omega)\right)$ is therefore governed primarily by magnetic correlations and phonons. Figure~\ref{fig:tr1} shows representative Raman spectra measured over a wide energy range from 40 to 1500~cm$^{-1}$ at 11, 100, 200, and 300~K (see Figs.~S6 and S7 in the SI for Raman spectra at all measured temperatures). \par

    TFO crystallizes in the orthorhombic \textit{Pbnm} space group, corresponding to the $D_{2h}$ point group \cite{Dubrovin2024}. At the Brillouin-zone center ($\Gamma$), 24 Raman-active phonon modes are expected: $7A_g \,\oplus\, 7B_{1g} \,\oplus\, 5B_{2g} \,\oplus\, 5B_{3g}$.
    
    Phonon mode symmetry and the corresponding assignments have been extensively studied in earlier Raman investigations of isostructural RFeO$_3$ compounds \cite{Weber2016PRB} and in single-crystalline TbFeO$_3$ \cite{Dubrovin2024}. In the spectral range below 750~cm$^{-1}$, the Raman response is dominated by first-order phonon modes. Analysis of the spectra reveals six of the seven expected $A_g$ modes, five of seven expected $B_{1g}$ modes, one of the five $B_{2g}$ modes, and two of the five $B_{3g}$ modes. The remaining modes are not clearly resolved at room temperature, primarily due to the limited spectral resolution ($2.5$\,cm$^{-1}$) and mode overlap. The observed phonon modes and their symmetry assignments are indicated in Fig.~\ref{fig:tr1}.
    
   The room-temperature phonon frequencies and the corresponding full widths at half maximum (FWHM), extracted from the Voigt-profile fits, are summarized in Table~\ref{tab:raman_exp_comp}. These values are compared with the available literature data for single-crystalline TbFeO$_3$. It should be noted that most single-crystal Raman studies employ polarization-dependent scattering geometries, which allow a more direct symmetry-resolved assignment of the $A_g$, $B_{1g}$, $B_{2g}$, and $B_{3g}$ modes. In contrast, the present measurements were performed on a polycrystalline specimen, where the measured Raman response represents an orientational average over different crystallographic grains. Therefore, the comparison in Table~\ref{tab:raman_exp_comp} is used primarily to establish the correspondence between the observed phonon frequencies and the reported Raman-active modes. Note that the linewidths for $B_{1g}$ (156~cm$^{-1}$) and $A_g$ (329~cm$^{-1}$) modes obtained here should be regarded as effective linewidths, containing both intrinsic phonon lifetime broadening and extrinsic inhomogeneous contributions arising from grain boundaries, microstrain, finite crystallite size, and local structural disorder with instrumental resolution.

\newcommand{\nd}{\multicolumn{1}{c}{--}}   
\begin{table*}[htbp]
    \centering
    \begin{threeparttable}
    \caption{Comparison of room-temperature Raman-active phonon modes of TbFeO$_3$  with reported values. All frequencies and linewidths are given in cm$^{-1}$. For the $B_{1g}$ (156~cm$^{-1}$) and $A_g$ (329~cm$^{-1}$) modes, which exhibit Gaussian-dominated line shapes, the total FWHM obtained from the Thompson-Cox-Hastings (TCH) formalism is listed. For all other modes, the FWHM corresponds to the Lorentzian component.}
    \label{tab:raman_exp_comp}
    \setlength{\tabcolsep}{7pt}
    \renewcommand{\arraystretch}{1.15}
    \providecommand{\nd}{\multicolumn{1}{c}{--}}

    \begin{tabular}{@{}lcccccc@{}}
        \toprule
        & \multicolumn{2}{c}{\textbf{Present work, 300 K}}
        & \multicolumn{4}{c}{\textbf{Reported experiment}} \\
        \cmidrule(lr){2-3} \cmidrule(lr){4-7}
        \textbf{Sym.}
        & \textbf{Freq.}
        & \textbf{FWHM}
        & \textbf{Freq.} \cite{Dubrovin2024}
        & \textbf{FWHM} \cite{Dubrovin2024}
        & \textbf{Freq.} \cite{Venugopalan1985}
        & \textbf{Freq.} \cite{Weber2016PRB} \\
        \midrule

        \multirow{7}{*}{$A_g$}
        & \nd & \nd & 111.0 & 3.3  & 109 & 112.5 \\
        & $135.42 \pm 0.21$ & $6.61 \pm 0.59$ & 139.7 & 6.5  & 140 & 143.9 \\
        & $257.15 \pm 6.18$ & $13.60 \pm 3.09$ & 257.9 & 14.0 & 273 & 261.9 \\
        & $328.99 \pm 0.02$ & $11.35 \pm 1.27$ & 331.6 & 7.4  & 329 & 334.5 \\
        & $403.17 \pm 1.32$ & $9.20 \pm 4.40$ & 407.8 & 9.9  & \nd & 410.9 \\
        & $408.69 \pm 4.05$ & $14.35 \pm 8.12$ & 415.4 & 25.3 & 406 & 420.1 \\
        & $483.36 \pm 0.04$ & $14.75 \pm 0.22$ & 487.2 & 10.8 & 480 & 490.1 \\
        \midrule
	
        \multirow{7}{*}{$B_{1g}$}
        & $105.72 \pm 0.34$ & $2.89 \pm 2.18$ & 109.8 & 5.6  & \nd & 107.7 \\
        & $155.80 \pm 0.61$ & $9.39 \pm 1.08$ & 159.1 & 6.8  & 139 & 160.1 \\
        & \nd & \nd & 296.6 & 14.0 & \nd & 302.7 \\
        & $354.05 \pm 0.22$ & $12.74 \pm 1.92 $& 359.0 & 9.9  & 329 & \nd \\
        & \nd & \nd & 483.6 & 9.4  & 479 & 485.6 \\
        & $532.84 \pm 1.94$ & $39.99 \pm 4.96$ & \nd & \nd & \nd & 535.8 \\
        & $636.10 \pm 1.17$ & $39.73 \pm 1.32$ & 639.6 & \nd & \nd & \nd \\
        \midrule

        \multirow{5}{*}{$B_{2g}$}
        & \nd & \nd & 128.1 & 6.2  & \nd & \nd \\
        & \nd & \nd & 321.5 & 12.2 & \nd & \nd \\
        & \nd & \nd & 428.1 & 20.2 & 418 & 433.3 \\
        & $463.64 \pm 0.29$ & $14.24 \pm 1.86$ & 466.8 & \nd & \nd & 468.8 \\
        & \nd & \nd & \nd & \nd & \nd & \nd \\
        \midrule

        \multirow{5}{*}{$B_{3g}$}
        & \nd & \nd & 149.1 & 5.2  & 159 & \nd \\
        & $251.76 \pm 2.41$ & $16.49 \pm 2.35$ & 252.4 & 17.7 & 249 & 251.9 \\
        & \nd & \nd & 356.2 & 10.8 & 354 & 359.2 \\
        & \nd & \nd & 426.0 & 20.7 & 426 & 427.7 \\
        & $622.97 \pm 0.12$ & $19.60 \pm 0.65$ & 629.3 & 69.1 & \nd & \nd \\

        \bottomrule
    \end{tabular}
    \end{threeparttable}
\end{table*}
    Here, we focus on a few representative modes that are clearly resolved, as labeled in Fig.~\ref{fig:tr1}. The $B_{1g}$ mode near 156~cm$^{-1}$ is associated with an out-of-phase vibration of the Tb ions along crystallographic $a$-axis. The $A_g$ mode near $329~\mathrm{cm^{-1}}$ originates from in-plane vibrations of the equatorial O-atoms within the $x$-$z$ plane. The $B_{1g}$ mode near 354~cm$^{-1}$ is related to collective oxygen vibrations, which lead to out-of-phase rotation of FeO$_6$ octahedra. At higher frequencies, the $A_g$ mode near 484~cm$^{-1}$ is associated with O (equatorial)-Fe-O(apical) bond bending vibrations. The $B_{3g}$ and $B_{1g}$ modes observed near 623~cm$^{-1}$ and 636~cm$^{-1}$, respectively, are attributed to in-phase Fe-O (equatorial) stretching and FeO$_6$ breathing vibrations. These room-temperature assignments follow earlier reports on TbFeO$_3$ and related orthoferrites \cite{Dubrovin2024,Weber2016PRB}.

    \textit{Two-magnon Raman scattering.} Having briefly assigned the first-order phonon modes associated with lattice vibrations, we now focus on the higher-energy Raman response beyond $\sim 750\,\mathrm{cm^{-1}}$. The most intense feature in this range is a broad, double-peak-like structure centered near $1250\,\mathrm{cm^{-1}}$ with a linewidth of $\sim 84\,\mathrm{cm^{-1}}$ at 11~K. Its peak position and linewidth are nearly temperature independent, while the intensity increases upon cooling. Such a broad response is inconsistent with a first-order phonon mode. Indeed, first-principles phonon calculations for TbFeO$_3$ place the upper cutoff for first-order phonons at $\sim 650\,\mathrm{cm^{-1}}$ \cite{Weber2016PRB}. Similar high-energy features reported in isostructural YFeO$_3$ were attributed to second-order phonon scattering dominated by the phonon density of states \cite{Ponosov2020}. Since this contribution does not bear direct relevance to magnetic correlations, we do not discuss it further. {\par}
    
    We now focus on the two weak but well-defined broad features that are observed at $11\,\mathrm{K}$, centered at approximately  $960\,\mathrm{cm^{-1}}$ and $1108\,\mathrm{cm^{-1}}$. These modes persist up to room temperature, with their intensities progressively enhanced upon cooling, while their substantial linewidths rule out a first-order origin. With the help of the following linear spin-wave theory calculations, we assign these features to two-magnon (2M) Raman scattering. 2M Raman scattering involves creation of pairs of magnons with opposite momenta. Thus, the Raman response reflects a weighted magnon density of states (MDOS) with an energy scale approximately twice that of single-magnon excitations \cite{Fleury1968,Kumawat2024}.{\par}
    
    \begin{figure}[htbp]
        \centering
        \includegraphics[width=1\linewidth]{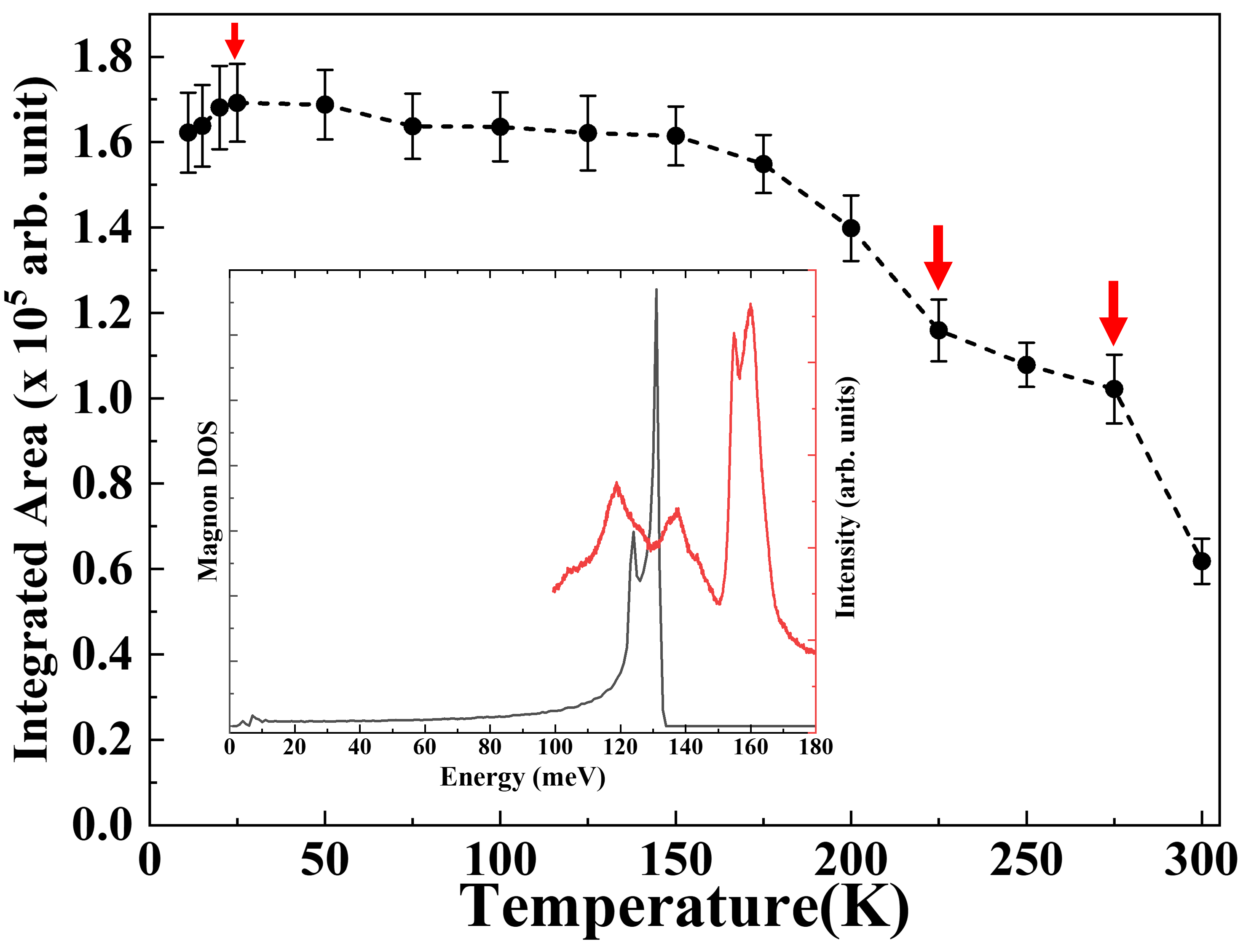}
        \caption{Temperature dependence of the integrated intensity of the two-magnon peak in polycrystalline TbFeO$_3$ powder. The inset shows the calculated magnon density of states (MDOS) for the Fe sublattice in TbFeO$_3$, plotted together with the two-magnon Raman spectrum measured at 11\,K for comparison.}
        \label{fig:trm1}
    \end{figure}
    
    We used linear spin-wave theory to calculate the MDOS. Since the lowest temperature of our Raman scattering experiments is above 10 K, i.e., above the ordering temperature of the Tb moments ($\sim 4$\,K), 2M Raman scattering is dominated by the Fe-sublattice. Further, we consider the ideal case, where the oxidation state of all Fe-ions is $+3$. The effective Hamiltonian for the Fe-sublattice includes isotropic exchange interactions, single-ion anisotropy, and the Dzyaloshinskii-Moriya interaction, which is written as
    \begin{equation}
        \begin{aligned}
        H^{\mathrm{Fe-Fe}} = &
        \sum_{ij} \mathbf{S}^{\mathrm{Fe}}_{i}\cdot J^{\mathrm{Fe}}_{ij}\cdot \mathbf{S}^{\mathrm{Fe}}_{j}
        + \sum_{i} \mathbf{S}^{\mathrm{Fe}}_{i}\cdot A^{\mathrm{Fe}}_{i}\cdot \mathbf{S}^{\mathrm{Fe}}_{i} \\
        & + \sum_{mn} \mathbf{S}^{\mathrm{Fe}}_{m}\cdot D^{\mathrm{Fe}}_{mn}\cdot \mathbf{S}^{\mathrm{Fe}}_{n}.
        \end{aligned}
        \label{eq:HH}
    \end{equation}
    Here $\mathbf{S}^{\mathrm{Fe}}_{i}$ denotes the spin operator for the Fe$^{3+}$ ion, $J^{\mathrm{Fe}}_{ij}$ represents the isotropic exchange interaction, $A^{\mathrm{Fe}}_{i}$ is the single-ion anisotropy tensor, and $D^{\mathrm{Fe}}_{mn}$ corresponds to the Dzyaloshinskii-Moriya interaction \cite{Ovsianikov2022}. 
    
    \begin{table}[t]
        \caption{Magnetic interaction parameters (in meV) used for TbFeO$_3$, adopted from \cite{Ovsianikov2022}}
        \label{tab:tbfeo3_parameters}
        \begin{ruledtabular}
             \begin{tabular}{ccccccc}
             $J_{ac}^{\mathrm{Fe}}$ & $J_{b}^{\mathrm{Fe}}$ & $J_{nnn}^{\mathrm{Fe}}$ & $D_1$ & $D_2$ & $A_{bc}$ & $A_a$ \\
            \hline
             4.77(1) & 4.55(2) & 0.10(2) & 0.13(3) & 0.10(3) & 0.007 & 0 \\
             \end{tabular}
        \end{ruledtabular}
    \end{table}
    
    The magnetic interaction parameters used in our calculations are summarized in Table \ref{tab:tbfeo3_parameters} for the high-temperature phase ($\Gamma_4$) of TFO. These values were taken from Ref.~\cite{Ovsianikov2022}, where these values were obtained by fitting inelastic neutron scattering data on single-crystal TbFeO$_3$ using the spin Hamiltonian given in Eq. \ref{eq:HH}.
    The lattice parameters of the sample studied in Ref. \cite{Ovsianikov2022} are very close to those obtained in the present work, as shown in Table \ref{tab:lattice_parameters}. This close structural agreement provides a reasonable physical basis for employing the reported exchange-interaction parameters in our 2MDOS calculations.

    Based on this Hamiltonian, using SpinW package~\cite{Toth2015}, we calculated MDOS and showed in the inset of Fig.~\ref{fig:trm1} by multiplying its energy-scale with $2$, as required due to simultaneous excitation of two magnons \cite{Kumawat2024}. The calculated MDOS shows pronounced contribution around $\sim1056\,\mathrm{cm^{-1}}$, which is in close proximity with the experimentally observed Raman modes at $1108\,\mathrm{cm^{-1}}$ and $960\,\mathrm{cm^{-1}}$. Thus, the origin of these modes are assigned to 2M Raman scattering. {\par}

    We examine the temperature evolution of the two-magnon (2M) Raman scattering by integrating the spectral weight over the range $801$-$1200$~cm$^{-1}$ around the modes. The resulting integrated spectral weight $(W_{2M})$ is shown in Fig.~\ref{fig:trm1}.The 2M scattering response persists throughout the measured temperature range of $11$ to $300\,\mathrm{K}$. This is consistent with the robust AF phase of TFO in this temperature range as $T_N \approx 650\,\mathrm{K}$. $W_{2M}$ increases upon cooling from 300\,K and becomes nearly temperature-independent below 150\,K. However, the overall evolution is non-monotonic; for instance, $W_{2M}$ exhibits a slight anomaly around 275\,K and slightly decreases below 25\,K. {\par}
    
    Within the Fleury-Loudon framework, the effective two-magnon light-scattering operator is written as
     \begin{equation}
        \hat O = \sum_{\langle ij\rangle} \eta_{ij}\,(\mathbf E_i\!\cdot\!\mathbf d_{ij})(\mathbf E_s\!\cdot\!\mathbf      d_{ij})\,\mathbf S_i\!\cdot\!\mathbf S_j ,
    \end{equation}
    where $\mathbf E_i$ and $\mathbf E_s$ are the polarization vectors of the incident and scattered photons, $\mathbf d_{ij}$ denotes the bond vector connecting spins $\mathbf S_i$ and $\mathbf S_j$, and $\eta_{ij}$ is the exchange Raman vertex, which scales with the superexchange interaction $J_{ij}$ \cite{Fleury1968,Sen2019}. In unpolarized measurements on a polycrystalline sample, the polarization projection factor $(\mathbf E_i\!\cdot\!\mathbf d_{ij})(\mathbf E_s\!\cdot\!\mathbf d_{ij})$ is largely orientation averaged. Thus, the dominant temperature dependence of the two-magnon intensity, i.e., $W_{2M}$ arises from terms proportional to $|\eta_{ij}|^2\langle(\mathbf S_i\!\cdot\!\mathbf S_j)^2\rangle$.  Here, $\langle \mathbf S_i\!\cdot\!\mathbf S_j\rangle$ represents spin-spin correlations, which is Fe-Fe spin correlations in the present context.  {\par}    
    The pronounced increase in DC magnetization below $\sim 50\,\mathrm{K}$ in Fig.~\ref{fig:mt} arises primarily from the paramagnetic response of the Tb moments \cite{Artyukhin2012}. Notably, this strong variation is not reflected in $W_{2M}$. Nevertheless, the Tb moments are polarized by the internal molecular field generated by the Fe sublattice \cite{Yamaguchi1974}, and the resulting Tb-Fe interaction may weakly influence the Fe-Fe spin correlations.{\par}
    Additionally, the anomalous variation in $W_{2M}$ near 275\,K coincides with the temperature at which the DC magnetization exhibits a shallow dip. This correlation suggests a subtle modification of the exchange interactions or the underlying spin correlations. Furthermore, the observation of two-magnon excitations, despite a positive Curie-Weiss temperature ($\theta_{\mathrm{CW}} > 0$), confirms the dominance of antiferromagnetic correlations, consistent with the intrinsic AF ground state of TFO.{\par}

\subsection{Non-trivial phonon renormalization}
\begin{figure}[htbp]
        \centering
        \includegraphics[width=1\linewidth]{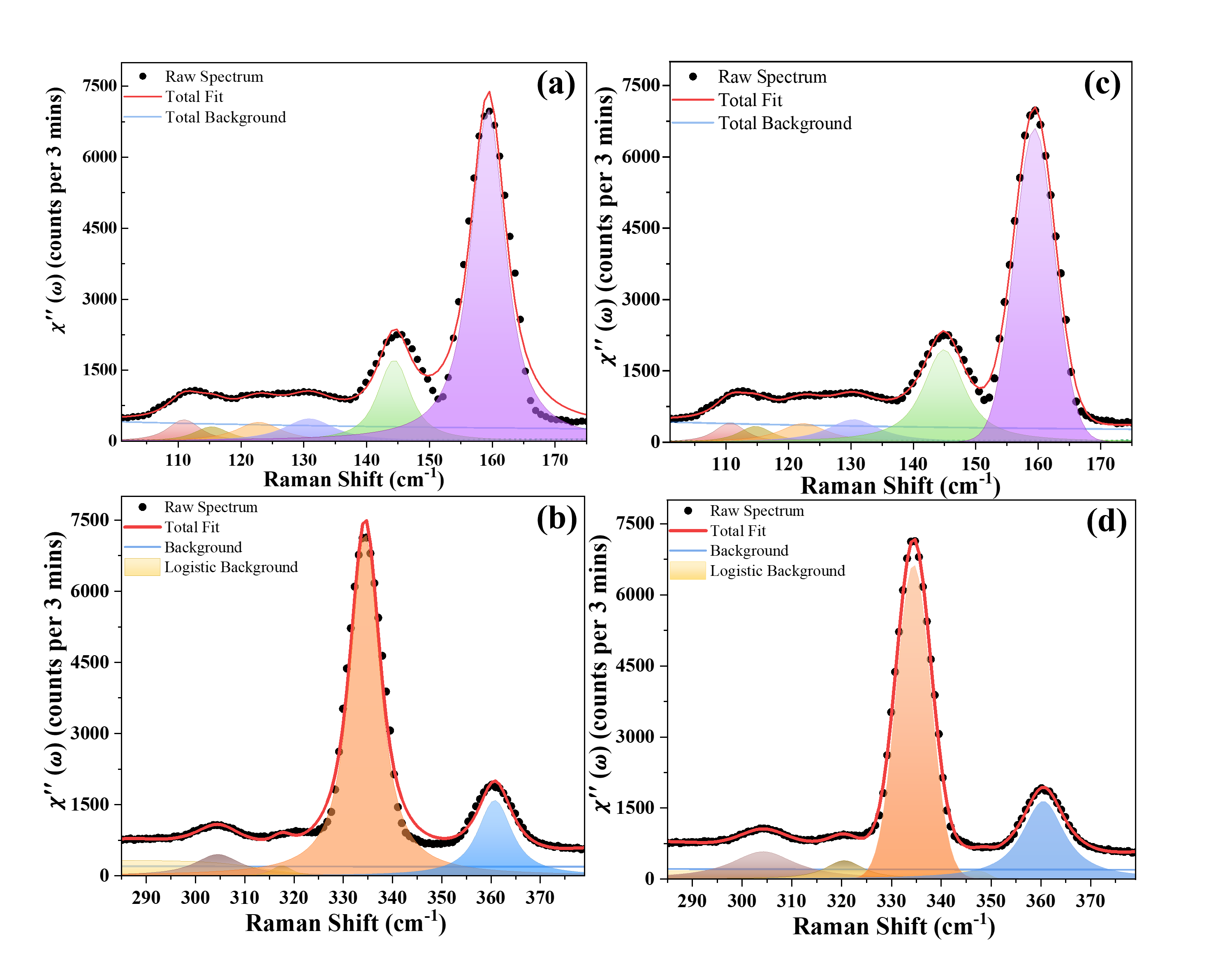}
        \caption{Voigt-function fits to the $B_{1g}$ (~156 cm$^{-1}$) and $A_g$ (~329 cm$^{-1}$) phonon modes measured at 11~K. The spectra are fitted under two conditions: (i) with the instrumental response function (IRF) fixed at 2.5~cm$^{-1}$ in the left column and (ii) with the IRF constraint released during the fitting procedure in the right column. The improved fit obtained without constraining the IRF suggests the presence of additional broadening beyond the instrumental resolution, possibly arising from local inhomogeneity in the material environment.}
    \label{fig:inh1}
    \end{figure}

    \textit{Phonon line-shape analysis.} We identify anomalous lattice dynamics by examining the temperature dependence of the phonon frequency ($\omega$) and linewidth ($\Gamma$). As noted earlier, the phonon modes are fitted using Voigt profiles, which account for the intrinsic Lorentzian response associated with a well-defined phonon frequency and finite lifetime, convolved with a Gaussian component corresponding to the instrumental resolution of 2.5\,cm$^{-1}$. 
    Among the observed modes, two prominent features at     $156~\mathrm{cm}^{-1}$ ($B_{1g}$) and $329~\mathrm{cm}^{-1}$ ($A_g$) exhibit clear deviations from a resolution-limited Lorentzian profile. A Voigt profile constrained by the instrumental Gaussian width of 2.5\,cm$^{-1}$ does not adequately reproduce the experimental spectra at all temperatures, as illustrated at 11\,K in Figs.~\ref{fig:inh1}(a) and (b). Satisfactory fits are obtained only when the Gaussian contribution is allowed to exceed the instrumental resolution, as shown in Figs.~\ref{fig:inh1}(c) and (d). At low temperatures, the best fits correspond to nearly pure Gaussian profiles, indicating that the linewidth is dominated by a distribution of phonon frequencies $(\omega_0 \rightarrow \omega_0 + \delta\omega)$ rather than a finite lifetime.

\begin{figure}[htbp]
        \centering
        \includegraphics[width=1\linewidth]{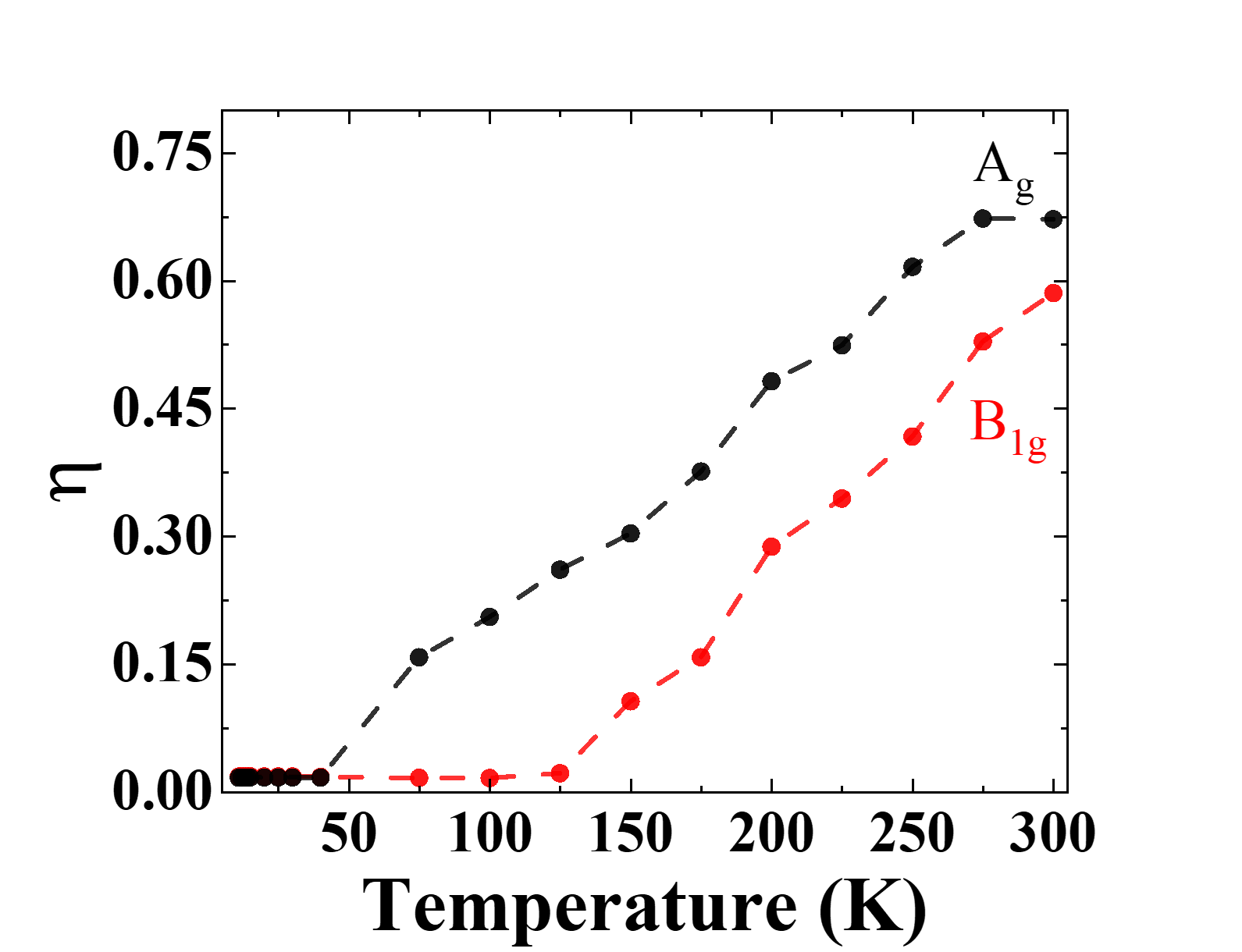}
        \caption{Temperature dependence of the Pseudo-Voigt shape parameter for the $B_{1g}$ (~156 cm$^{-1}$) and $A_g$ (~329 cm$^{-1}$) phonon modes. The evolution of the shape parameter reflects the relative contributions of Gaussian and Lorentzian broadening components. For the $A_g$ mode, the Lorentzian contribution becomes negligible below $\sim$50~K, while for the $B_{1g}$ mode a similar suppression of the Lorentzian component is observed below $\sim$125~K, indicating that the line shape in these temperature ranges is dominated by Gaussian broadening.}
        \label{fig:inh3}
    \end{figure}

    To quantify the relative contributions of Lorentzian $(L)$ and Gaussian $(G)$ components, we employ the Thomas-Cox-Hastings (TCH) formalism~\cite{Thompson1987}, writing the line shape as $I(\omega)=\eta\,L(\omega)+(1-\eta)\,G(\omega)$, where $\eta$ ($0\leq\eta\leq1$) denotes the Lorentzian weight. In practice, $\eta$ cannot reach unity due to the finite instrumental resolution. For both modes, $\eta$ increases approximately linearly with temperature up to 300\,K, indicating a progressive enhancement of the Lorentzian (lifetime) contribution, as shown in Fig.~\ref{fig:inh3}. The crossover from Gaussian-dominated to mixed line shape occurs above $\sim 40$\,K for the $A_g$ mode and above $\sim 125$\,K for the $B_{1g}$ mode. Overall, the Gaussian profile is a low-temperature characteristic, while the Lorentzian contribution becomes increasingly significant at higher temperatures.

    Within the Raman scattering framework, a Lorentzian line shape reflects homogeneous broadening arising from a finite phonon lifetime, typically governed by anharmonic phonon-phonon interactions. In contrast, a Gaussian contribution indicates inhomogeneous broadening associated with a distribution of local phonon frequencies \cite{Johnson2011}. In \cite{Johnson2011} it is attributed to short-range structural disorder in the crystal lattice, likely arising from angular distortions of the tetrahedral bonds. In TbFeO$_3$, the dominance of the Gaussian component at low temperatures indicates that the linewidth is governed primarily by a distribution of local phonon frequencies, $\omega_0 \rightarrow \omega_0 + \delta\omega$, arising from static variations in the lattice environment.
    Such variations can originate from static local inhomogeneity in the crystal. In the present polycrystalline sample, oxygen non-stoichiometry revealed by XPS, together with grain boundaries, microstrain, and finite crystallite size, can generate spatial variations in the local lattice environment and hence a distribution of local phonon frequencies. Since these structural inhomogeneities remain essentially unchanged over the investigated temperature range, they are expected to provide a largely temperature-independent contribution to the linewidth, which is accounted for by the residual linewidth term $\Gamma_{0}$ in the Klemens analysis. Therefore, they cannot by themselves explain the pronounced temperature evolution of the Gaussian component. We therefore consider that the low-temperature enhancement of the Gaussian contribution is more likely associated with the evolution of local magnetic inhomogeneity through spin-phonon coupling, although oxygen non-stoichiometry may indirectly contribute by modifying the local exchange interactions. This leads to a spread in the equilibrium positions and hence the phonon energies across the sample. In this regime, the intrinsic phonon lifetime is comparatively long, and homogeneous (Lorentzian) broadening remains weak.

    With increasing temperature, anharmonic phonon-phonon interactions become significant and introduce a finite phonon lifetime, giving rise to a Lorentzian contribution to the line shape. Consequently, the observed profile evolves into a mixed (Gaussian + Lorentzian) form, reflecting the coexistence of static inhomogeneous broadening and temperature-dependent lifetime effects. The crossover from Gaussian-dominated to mixed line shapes therefore captures the transition from a regime governed by spatial variations in $\omega_0$ to one where anharmonic decay processes play an increasingly important role in determining the Raman response. The temperature evolution of the shape parameter reveals distinct onsets at $\sim$50~K for the $A_g$ mode and $\sim$125~K for the $B_{1g}$ mode. These trends exhibit a consistent correspondence with the temperature-dependent XRD results. The $A_g$ mode, primarily associated with oxygen vibrations in the $x$-$z$ ($a$-$c$) plane, shows an onset near $\sim$50~K, which coincides with the temperature range where the $c$ lattice parameter remains nearly constant. Similarly, the $B_{1g}$ mode, involving Tb atomic displacements along the $a$ direction, gaussian nature aligns with the nearly temperature-independent behavior of the $a$ lattice parameter below $\sim$125~K, as shown in Fig.~\ref{fig:xrd}(a). This correspondence indicates a close interplay between lattice distortions and phonon dynamics. \par

\begin{figure}[htbp]
          \centering
          \includegraphics[width=1\linewidth]{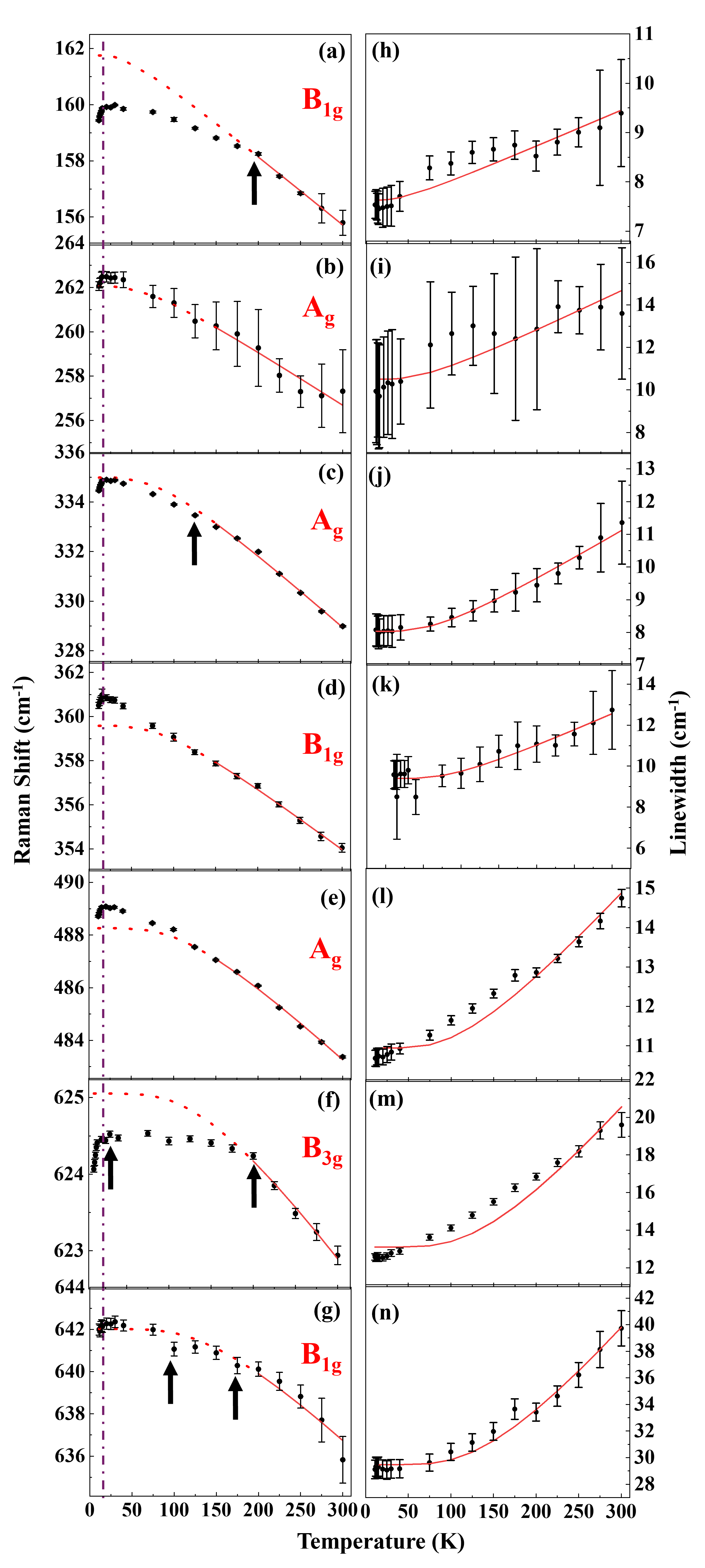}
          \caption{Temperature evolution of selected Raman-active phonon modes analyzed within the Klemens anharmonic decay model. Panels (a–g) present the temperature dependence of the phonon mode center (Raman shift) for different modes, while panels (h–n) shows the corresponding phonon linewidths (FWHM) as a function of temperature. The symbols represent the experimentally extracted phonon parameters obtained from Voigt-profile fitting, and the associated error bars correspond to the fitting uncertainties; in several cases the error bars are smaller than the symbol size and are therefore not readily visible. Solid red lines represent fits to the Klemens anharmonic model, capturing the temperature-driven renormalization of phonon energies and lifetimes arising from multi-phonon decay processes. Red dotted lines in the Raman shift panels indicate extrapolations of the Klemens fit beyond the temperature range over which the model adequately describes the experimental data.}
        \label{fig:kle1}
    \end{figure}

    \textit{Lattice dynamics beyond conventional lattice anharmonicity.} Figs.~\ref{fig:kle1}(a)–(g) and \ref{fig:kle1}(h)–(n) show the temperature dependence of the phonon frequencies ($\omega$) and the corresponding linewidths ($\Gamma$) of the prominent Raman-active modes. Among these, the two modes exhibiting Gaussian-dominated line shapes at low temperature, namely $B_{1g}$ ($\sim 156~\mathrm{cm}^{-1}$) and $A_g$ ($\sim 329~\mathrm{cm}^{-1}$), are highlighted in Figs.~\ref{fig:kle1}(a),(h) and \ref{fig:kle1}(c),(j), respectively. The remaining modes are well described by Lorentzian line shapes over the entire temperature range. The objective here is to examine whether the phonons exhibit deviations from conventional lattice anharmonicity, thereby indicating coupling to other degrees of freedom.

    For a magnetic insulator, the temperature dependence of the phonon frequency can be expressed as~\cite{Granado1999}
    \begin{equation}
        \omega(T) = \omega(0) + \Delta\omega_{\mathrm{qh}}(T) + \Delta\omega_{\mathrm{anh}}(T) + \Delta\omega_{\mathrm{sp\!-\!ph}}(T),
    \end{equation}
    where $\omega(0)$ is the harmonic phonon frequency in the $T \rightarrow 0$ limit. The term $\Delta\omega_{\mathrm{qh}}(T)$ arises from quasi-harmonic effects associated with volume changes, $\Delta\omega_{\mathrm{anh}}(T)$ represents intrinsic anharmonic contributions due to phonon-phonon decay processes, and $\Delta\omega_{\mathrm{sp\!-\!ph}}(T)$ accounts for spin-phonon coupling. The latter can be written as
    \begin{equation}
        \Delta \omega_{sp\!-\!ph} = \lambda \langle \mathbf S_i \cdot \mathbf S_j \rangle,
    \end{equation}
    where $\langle \mathbf S_i \cdot \mathbf S_j \rangle$ is the spin-spin correlation function. The spin-phonon coupling constant $\lambda$ originates from the dependence of the exchange interaction $(J)$ on atomic displacement $(u)$ and is proportional to $\partial^2 J/\partial u^2$.

    In TbFeO$_3$, the relative change in unit-cell volume over the measured temperature range is extremely small ($\sim 0.017\%$). Therefore, the quasi-harmonic contribution $\Delta\omega_{\mathrm{qh}}(T)$ is negligible, and the temperature dependence of $\omega(T)$ is primarily governed by anharmonic and spin-phonon contributions.

    The intrinsic anharmonic behavior is described by the Klemens decay model, in which an optical phonon decays into two acoustic phonons of equal and opposite momenta~\cite{Klemens1966,MenendezCardona1984}. The corresponding expressions for $\omega(T)$ and $\Gamma(T)$ are
    \begin{equation}
        \omega(T) = \omega_{0} - C\left( 1 + \frac{2}{e^{\hbar \omega_{0}/2k_{\mathrm{B}}T} - 1} \right),
        \label{eq:klemens_freq}
    \end{equation}
    \begin{equation}
        \Gamma(T) = \Gamma_{0} + \Gamma\left( 1 + \frac{2}{e^{\hbar \omega_{0}/2k_{\mathrm{B}}T} - 1} \right),
        \label{eq:klemens_lw}
    \end{equation}
    where $\omega_{0}$ is the bare phonon frequency, $C$ is the anharmonic constant, $\Gamma_{0}$ represents the residual linewidth arising from defects, and $\Gamma$ is the anharmonic contribution.

    The high-temperature region of the data is fitted using the Klemens model, and the extrapolated behavior is shown by the dotted lines in Fig.~\ref{fig:kle1}. As expected, the model captures the general trend: $\omega(T)$ increases and saturates at low temperature, while $\Gamma(T)$ decreases and approaches a residual value $\Gamma_0$.

    However, clear deviations from this behavior are observed for several modes. In particular, the $B_{1g}$ mode near $156~\mathrm{cm}^{-1}$ and the $B_{3g}$ mode near $623~\mathrm{cm}^{-1}$ exhibit anomalous softening upon cooling, whereas the $B_{1g}$ mode near $354~\mathrm{cm}^{-1}$ and the $A_g$ mode near $484~\mathrm{cm}^{-1}$ show enhanced hardening relative to the Klemens expectation. Earlier, deviations of the phonon frequencies from the Klemens anharmonic model have also been reported for TbFeO$_3$~\cite{Vilarinho2022}. Since $\langle \mathbf S_i \cdot \mathbf S_j \rangle$ is a bulk quantity common to all modes, the sign and magnitude of these deviations are governed by the mode-dependent coupling constant $\lambda$. In this regard the present work provides a qualitative discussion of spin-phonon coupling rather than a quantitative determination of the coupling strength. The mode-resolved estimation of $\lambda$ would require an independent determination of the magnetic correlation function and is beyond the scope of the present work. Notably, these anomalies correlate with features in the Fig.~\ref{fig:trm1}, indicating a coupling between lattice and magnetic degrees of freedom. 
    At low temperatures ($\lesssim 20$\,K), most modes exhibit additional softening. This temperature range coincides with the onset of strong correlations in the Tb sublattice and its interaction with the Fe sublattice. Such correlations can modify the exchange interactions and thereby influence the phonon energies through spin-phonon coupling.

    Overall, these results demonstrate a clear coupling between lattice dynamics and magnetic correlations in TbFeO$_3$, going beyond conventional anharmonic phonon behavior.

    \textit{Emergence of a new mode below 175\,K} 
    Finally, we focus on a weak, broad mode that appears at low temperatures, as indicated by an arrow in Fig.~4. To track the temperature evolution of this mode, we evaluate the difference spectrum $\chi^{\prime\prime}(\omega,T)-\chi^{\prime\prime}(\omega,300\,\mathrm{K})$ and integrate it over the spectral window around the mode, as shown in Fig.~9(a).The corresponding integrated area is plotted as a function of temperature in Fig.~9(b). Clearly, the mode develops upon cooling with an onset around $\sim 175$\,K and becomes well-defined at low temperatures. The temperature dependence of the integrated area, $A(T)$, is reasonably described by a phenomenological power-law onset form, $A(T)\propto (T^{\ast}-T)^{\beta}$ \cite{Zhou2019}, yielding $T^{\ast}\approx175$\,K and $\beta\approx0.36$. Here, $\beta$ is treated only as an effective fitting parameter describing the gradual growth of the Raman spectral weight. The fit is employed solely to quantify the onset-like evolution of the emergent Raman response. At present, we are unable to identify a well-defined microscopic order parameter associated with this behavior.
     \begin{figure}[htbp]
        \centering
        \includegraphics[width=1\linewidth]{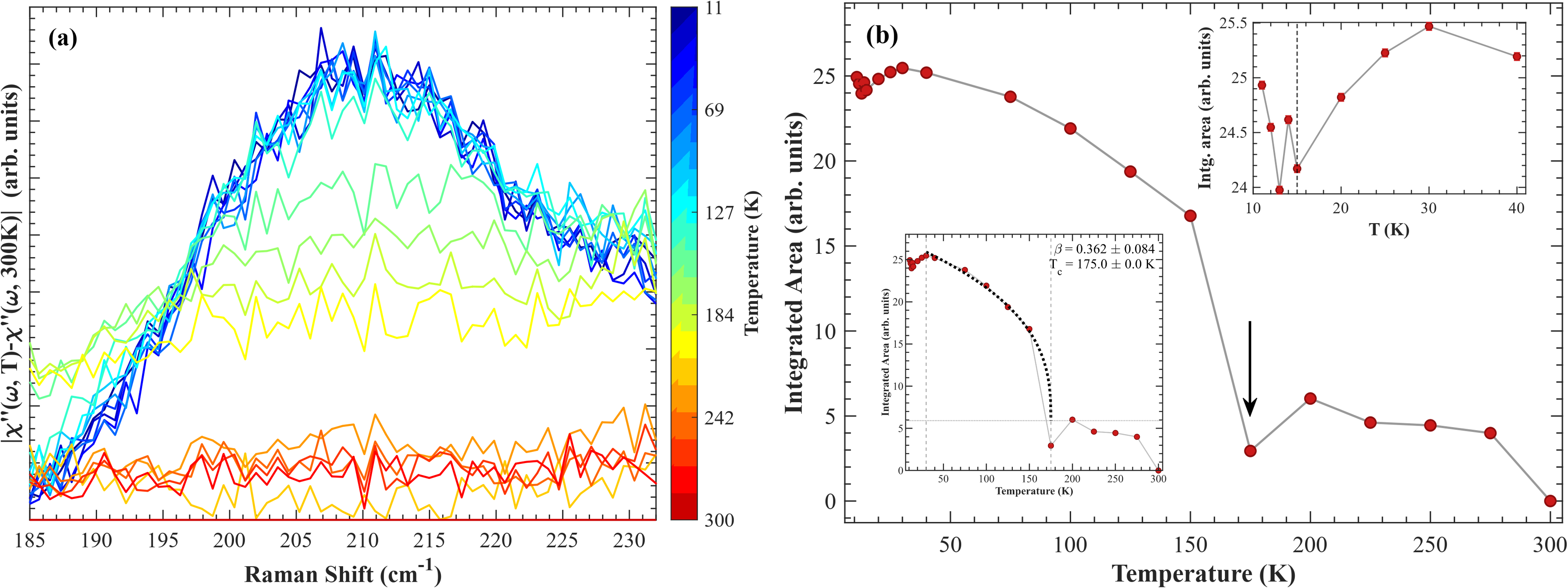}
        \caption{(a) Temperature dependence of the difference spectrum $|X(\omega,T)-X(\omega,300\,\mathrm{K})|$ for peak observed ~206 cm$^{-1}$, showing the evolution of the spectral response relative to the 300\,K reference. (b) Temperature dependence of the integrated area of the difference spectra. A clear change in trend is observed below $\sim175$\,K. The upper right inset highlights the low-temperature behavior in the 10-40\,K range, where an additional change is observed below $\sim15$\,K. The lower-left inset presents the phenomenological power-law fit, yielding an onset temperature $T^{\ast}\approx175$~K.}
        \label{fig:ordP}
    \end{figure}

    Notably, the onset temperature of the broad Raman feature near 206~cm$^{-1}$ coincides with the temperature range where several first-order phonons deviate from the symmetric anharmonic behavior described by the Klemens model (see Fig. 8). However, the integrated two-magnon spectral weight shows only a weak kink near this temperature, without any pronounced anomaly (see Fig.5). Since the Fe sublattice is already antiferromagnetically ordered below $T_N \sim 650$K, the anomaly near $\sim 175$K is unlikely to originate from a primary establishment of Fe magnetic order. Instead, the observations indicate that the emergence of the broad Raman feature is accompanied by modifications of the lattice dynamics, while the long-range Fe antiferromagnetic order remains largely unaffected. {\par} 
    
    This interpretation is further supported by the subtle change in the integrated response around $\sim 15$~K, as shown in the inset of Fig.~9(b). This temperature range coincides with the onset of additional phonon softening and a pronounced increase in magnetization associated with the Tb moments prior to their long-range antiferromagnetic ordering below $\sim 4$,K. A corresponding reduction in the integrated two-magnon spectral weight is also observed (see Fig.~5). Taken together, these observations suggest that the low-temperature Raman response is influenced by the evolving Tb-Fe magnetic correlations, and the accompanying changes in the local lattice environment. At present, however, the microscopic origin of the crossover near $\sim175$ K remains an open question. We note that $T^{\ast}$ lies close to the dielectric anomaly reported near 200~K in single-crystalline TbFeO$_3$ \cite{Indra2025}; however, the present data do not allow us to establish whether the two share a common origin.


\section{Summary and Conclusions}
In this work, we have carried out a comprehensive temperature-dependent investigation of polycrystalline TbFeO$_3$ using x-ray diffraction (XRD), x-ray photoelectron spectroscopy (XPS), DC magnetization, and Raman scattering. The combined measurements provide an extensive set of bulk properties of the material. {\par}

Temperature-dependent XRD measurements report, for the first time, the presence of negative thermal expansion over the range 5–300 K, with a small but measurable increase in unit-cell volume upon cooling and no evidence of any structural phase transition. The most evident anomaly occurs along the $c$ axis near $\sim175$~K, where the thermal response changes from contraction to expansion, coinciding with the emergence of a new Raman mode. XPS measurements reveal mixed-valence states of Fe and Tb, indicating non-stoichiometry and the presence of multiple local electronic environments. {\par}

Magnetization measurements confirm the antiferromagnetic ordering of the Fe sublattice near $T_N \sim 650$ K with weak ferromagnetic canting. In addition, the magnetization data reveal further features at lower temperatures associated with the complex interplay between the Fe and Tb sublattices. {\par}

Raman scattering measurements identify most of the first-order Raman-active phonon modes. The temperature evolution of many of these modes shows clear departures from the Klemens model, primarily in the phonon frequencies, indicating the presence of additional interactions beyond conventional phonon-phonon scattering. For two phonon modes of $A_g$ and $B_{1g}$ symmetry, we observe Gaussian-dominated line shapes at low temperatures instead of the expected Lorentzian response. With increasing temperature, these modes evolve toward mixed Gaussian-Lorentzian profiles. This behavior indicates that, at low temperatures, these phonons are affected predominantly by static local inhomogeneity, leading to a distribution of local phonon frequencies, $\omega_0 \rightarrow \omega_0 + \delta\omega$. {\par}

In addition, the high-energy Raman spectra reveal two-magnon excitations associated with the Fe sublattice. This assignment is supported by linear spin-wave theory calculations based on the reported exchange coupling constants for TbFeO$_3$. The temperature dependence of the integrated two-magnon spectral weight remains largely consistent with a robust antiferromagnetic background, with only weak anomalies at intermediate temperatures. {\par}

A broad Raman mode emerging below $\sim 175$ K exhibits an onset-like temperature dependence. Its onset coincides with the temperature range in which several phonon modes begin to deviate from conventional anharmonic behavior, while no corresponding strong anomaly is observed in the two-magnon response. This indicates that the feature is not associated with a primary magnetic ordering transition of the Fe sublattice. Rather, it points to a more subtle crossover involving the local spin-lattice environment. {\par}

Since the present study was performed on a polycrystalline sample, it is useful to distinguish intrinsic properties from those that may be affected by the microstructure. The lattice parameters and phonon frequencies agree with single-crystal data. The phonon linewidths, however, contain a temperature-independent contribution from grain boundaries, oxygen non-stoichiometry, while their temperature-dependent Gaussian broadening is attributed to the local magnetic inhomogeneity through spin-phonon coupling. In addition, orientational averaging prevents a symmetry-resolved analysis. {\par}

Overall, the combined structural, magnetic, and spectroscopic results demonstrate that TbFeO$_3$ hosts a pronounced interplay among lattice dynamics, spin correlations, and local magnetic environment over a wide temperature range, without undergoing any structural phase transition. More broadly, these results identify TbFeO$_3$ as a useful model system for studying spin-lattice coupling and correlated lattice anomalies in rare-earth orthoferrites and related complex oxides.

\begin{acknowledgments}
We acknowledge support from the Department of Physics and the Central Research Facilities at IIT Delhi. K.S. acknowledges the following research grants: INSPIRE Faculty Fellowship from DST, India; the Core Research Grant (CRG) from ANRF, India; and the YSRP grant from BRNS, India. S.F. acknowledges a PhD fellowship under the INSPIRE scheme from DST, India. DM acknowledges the funding from the India Russia Joint Research project, Department of Science and Technology, Government of India (Grant No. DST/IC/RSF/2024/542) and the Anusandhan National Research Foundation, Government of India (Grant No. ARNF/ARG/2025/007161/PS).
\end{acknowledgments}

\bibliographystyle{apsrev4-2}
\bibliography{refs}
\clearpage
\onecolumngrid
\appendix
\setcounter{figure}{0}
\renewcommand{\thefigure}{S\arabic{figure}}

\section{Supplementary Information}

\section{XRD}
\begin{figure}[htbp]
    \centering
    \includegraphics[width=0.5\textwidth]{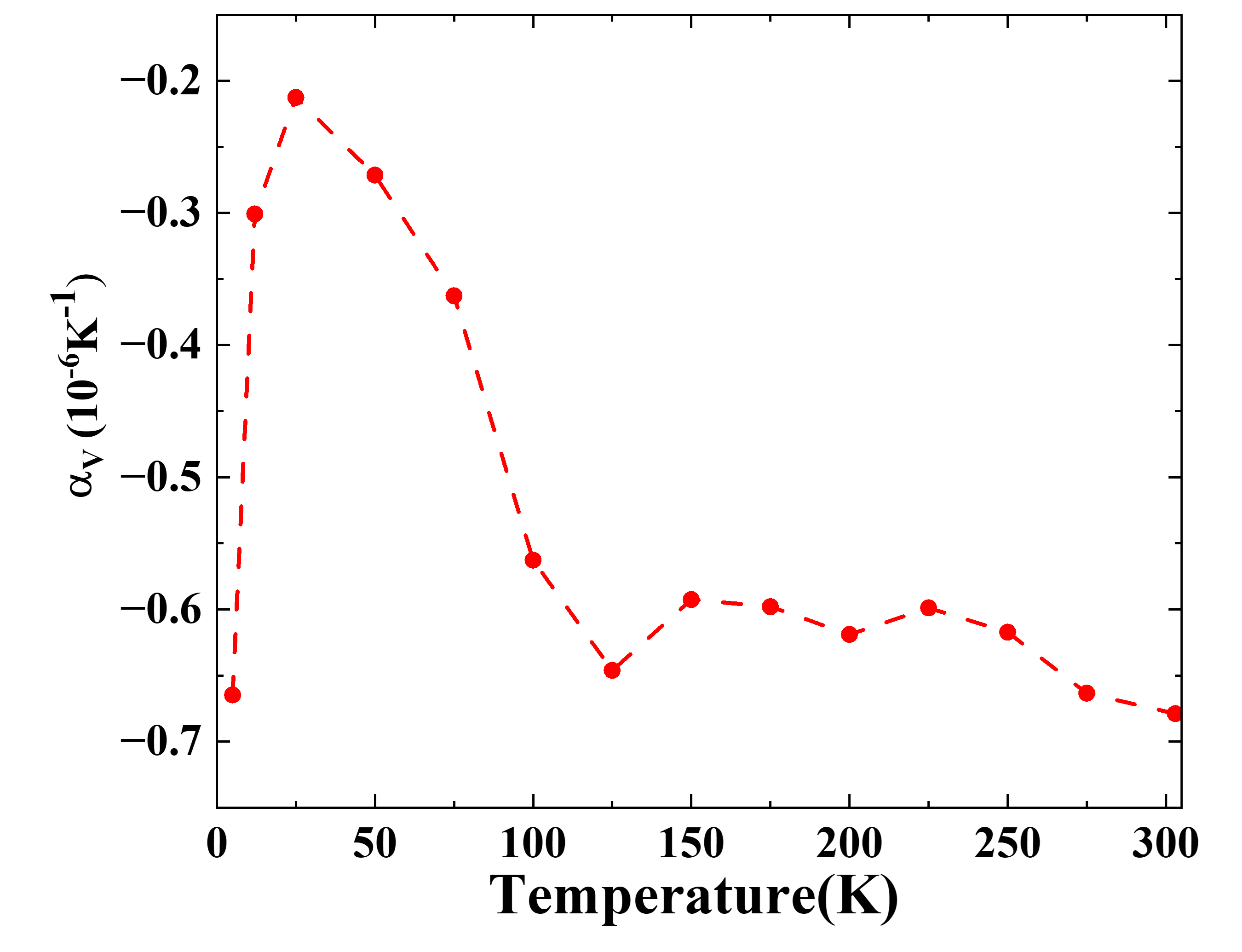}
    \caption{Temperature dependence of the instantaneous thermal expansion response.}
    \label{fig:thermal_expansion_2}
\end{figure}
\section{EDX}

Energy-dispersive X-ray spectroscopy (EDX) measurements were performed using a Hitachi FlexSEM 1000 II to determine the elemental composition of the sample. The EDX spectra confirm the presence of the constituent elements Tb, Fe, and O. Measurements were carried out at three different locations on the sample surface. The averaged atomic concentrations obtained from these measurements are Fe: 22.60\%, Tb: 22.84\%, and O: 54.55\%. During the analysis, contributions from Au and C peaks were excluded. The Au signal originates from the conductive gold coating applied for SEM measurements, while the C peak arises from the carbon tape used to mount the sample. After removing these extrinsic contributions, the elemental composition was recalculated using the remaining characteristic peaks of Tb, Fe, and O.A weak Mg signal is also observed in the spectrum. This feature likely arises from the overlap between the Mg K$\alpha$ emission line (1.253~keV) and the Tb M-shell emission line ($\sim$1.240~keV). The small energy difference ($\sim$0.013~keV) lies within the typical energy resolution of EDX detectors, making these peaks difficult to distinguish reliably. Therefore, the apparent Mg contribution is attributed to this spectral overlap rather than the presence of Mg in the sample.

    \begin{figure}[H]
        \centering
        \includegraphics[width=\textwidth]{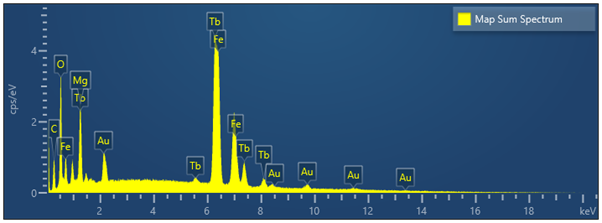}
        \caption{Raw energy-dispersive X-ray spectroscopy (EDX) spectrum of TbFeO$_3$. The spectrum shows characteristic peaks corresponding to Tb, Fe, and O. Additional signals from Au and C originate from the gold coating used for SEM measurements and the carbon tape used for sample mounting, respectively.}
        \label{fig:S1}
    \end{figure}

    \begin{figure}[H]
        \centering
        \includegraphics[width=\textwidth]{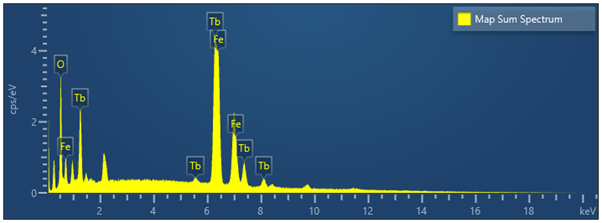}
        \caption{EDX spectrum used for elemental composition analysis after removing peaks from Au and C. The remaining peaks corresponding to Tb, Fe, and O were used to estimate the average atomic composition of the sample.}
        \label{fig:S2}
    \end{figure}

\section{XPS}
      The atomic fraction of element $p$ was calculated using the standard XPS quantitative relation
        \begin{equation}
            X = \frac{I_p/S_p}{\sum_j I_j/S_j},
        \end{equation}
    where $I_p$ is the integrated peak area of the photoelectron line of element $p$ and $S_p$ is the corresponding relative sensitivity factor. The summation runs over all detected elements in the spectrum \cite{Shard2020}. 
      \begin{figure}[H]
        \centering
        \includegraphics[width=\textwidth]{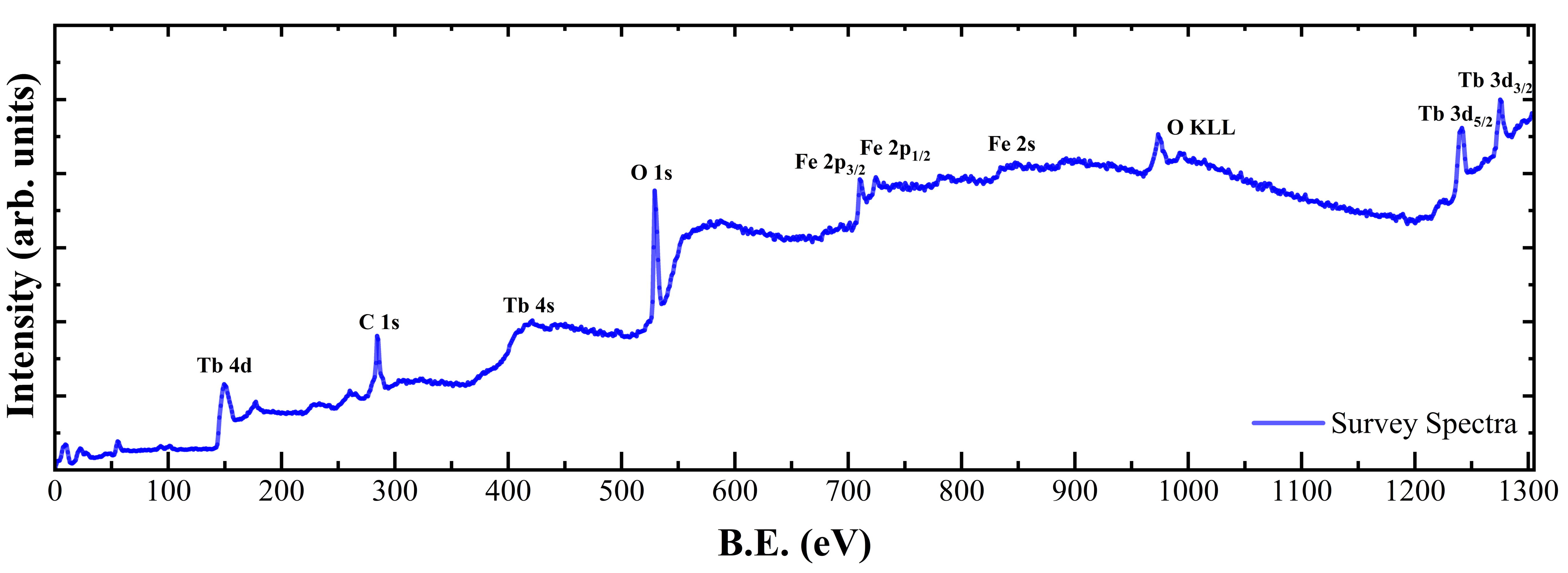}
        \caption{X-ray photoelectron spectroscopy (XPS) survey spectrum of TbFeO$_3$. The spectrum confirms the presence of the constituent elements Tb, Fe, and O without detectable impurity peaks. The characteristic core-level signals corresponding to Tb $3d$, Fe $2p$, and O $1s$ are clearly observed, verifying the elemental composition of the sample. The absence of additional peaks indicates phase purity within the detection limit of XPS.}
        \label{fig:S3}
    \end{figure}
    
\section{Magnetization}

    \begin{figure}[H]
        \centering
        \includegraphics[width=0.9\linewidth]{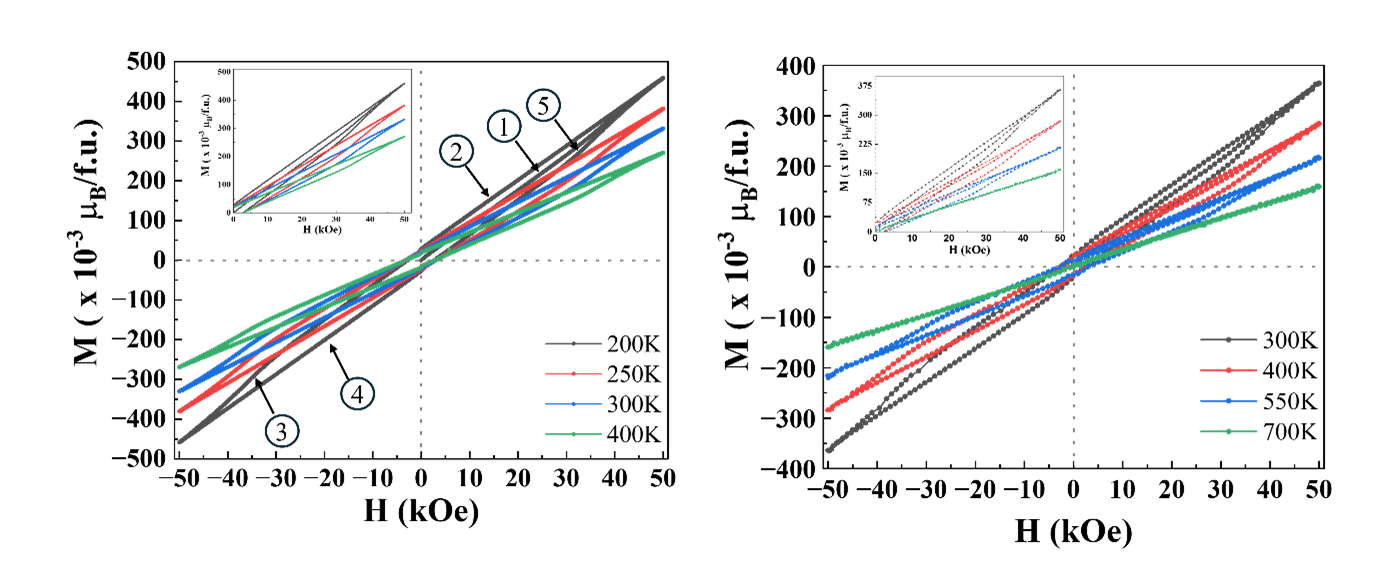}
        \caption{
            Isothermal field-dependent magnetization $M(H)$ of TbFeO$_3$. Fig.~S1(a) and S1(b) correspond to data obtained using two different measurement assemblies. Over the complete field cycle, the magnetization exhibits distinct behavior across five segments. At 200~K, the virgin branch shows a weak tendency toward slope enhancement at higher fields. This effect becomes significantly amplified in the third ($0 \rightarrow -5$~T) and fifth ($0 \rightarrow +5$~T) segments, where the $M(H)$ curve develops an increase in constant slope beyond a characteristic field. In contrast, the second and fourth branches, corresponding to the field returning from full positive or negative saturation, follow an almost perfectly linear descent from the saturated state, yielding well-defined remanent magnetization $M_r$ at $H=0$ and coercive fields $H_c$. This asymmetry indicates that repeated field cycling reshapes the magnetic domain landscape. At temperatures above the magnetic ordering temperature ($\sim650$~K), the $M(H)$ curves become strictly linear over the entire field range, consistent with paramagnetic behavior. This confirms that the anomalous field-dependent features observed at lower temperatures arise from the ferromagnetically ordered state. The temperature evolution of the hysteresis parameters further supports the weak ferromagnetic nature of TbFeO$_3$. At 200~K, the remanent magnetization is $M_r^{+} \approx 2.8 \times 10^{-3}\,\mu_B$/f.u.\ and $M_r^{-} \approx -2.7 \times 10^{-3}\,\mu_B$/f.u., while the coercive fields are $H_c^{-} \approx -2.8$~kOe and $H_c^{+} \approx 2.7$~kOe. With increasing temperature, both the remanent magnetization and coercivity decrease, reflecting a gradual weakening of the canted antiferromagnetic order. At higher temperatures ($\sim700$~K), the hysteresis nearly vanishes, consistent with the approach to the paramagnetic state.}
        \label{fig:S4}
    \end{figure}

\section{Raman Scattering}   
 
    \begin{figure}[H]
        \centering
        \includegraphics[width=\textwidth]{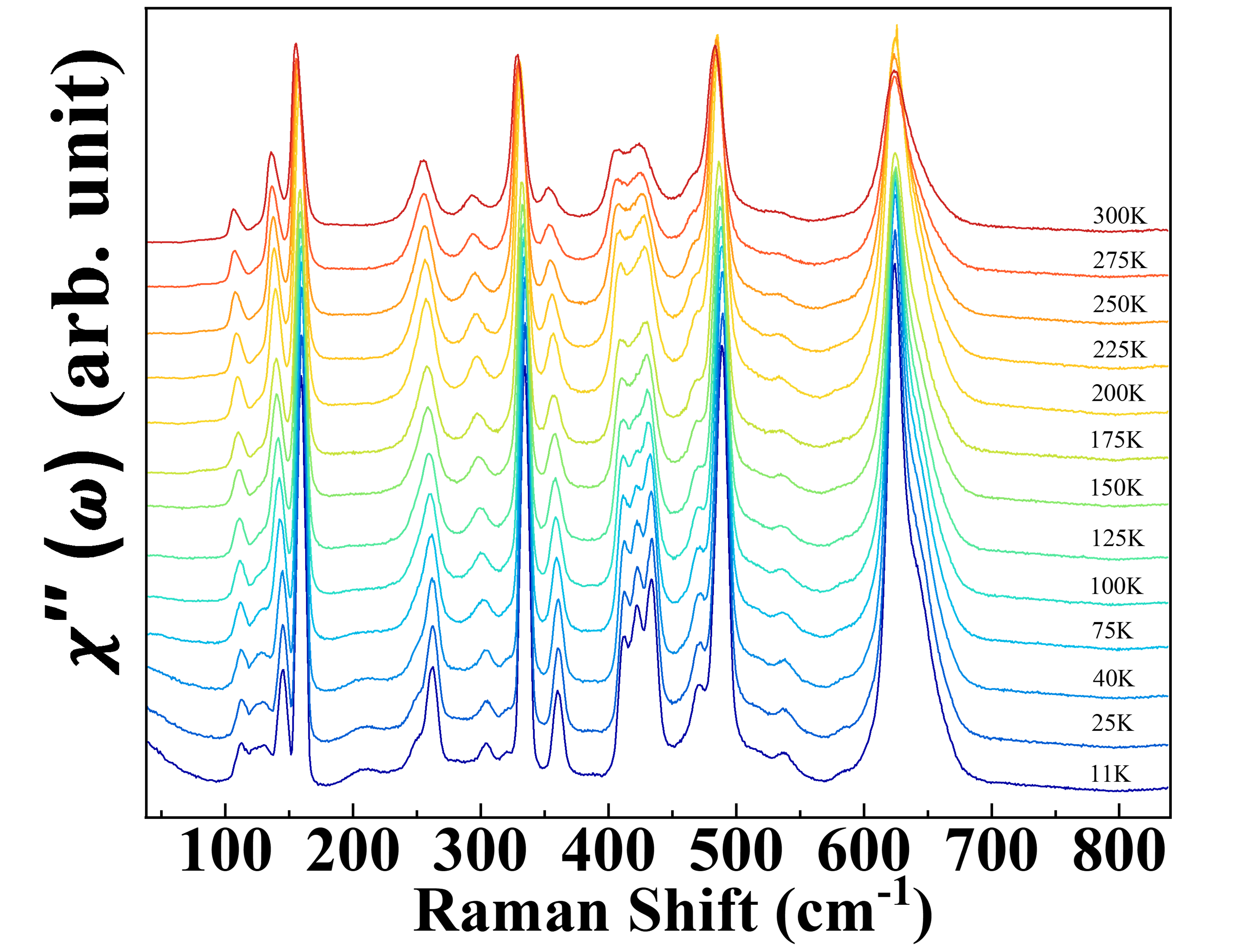}
        \caption{Temperature-dependent Raman scattering spectra of TbFeO$_3$ (TFO) showing the evolution of phonon modes in the low-wavenumber region.}
        \label{fig:S5}
    \end{figure}

        \begin{figure}[H]
        \centering
        \includegraphics[width=\textwidth]{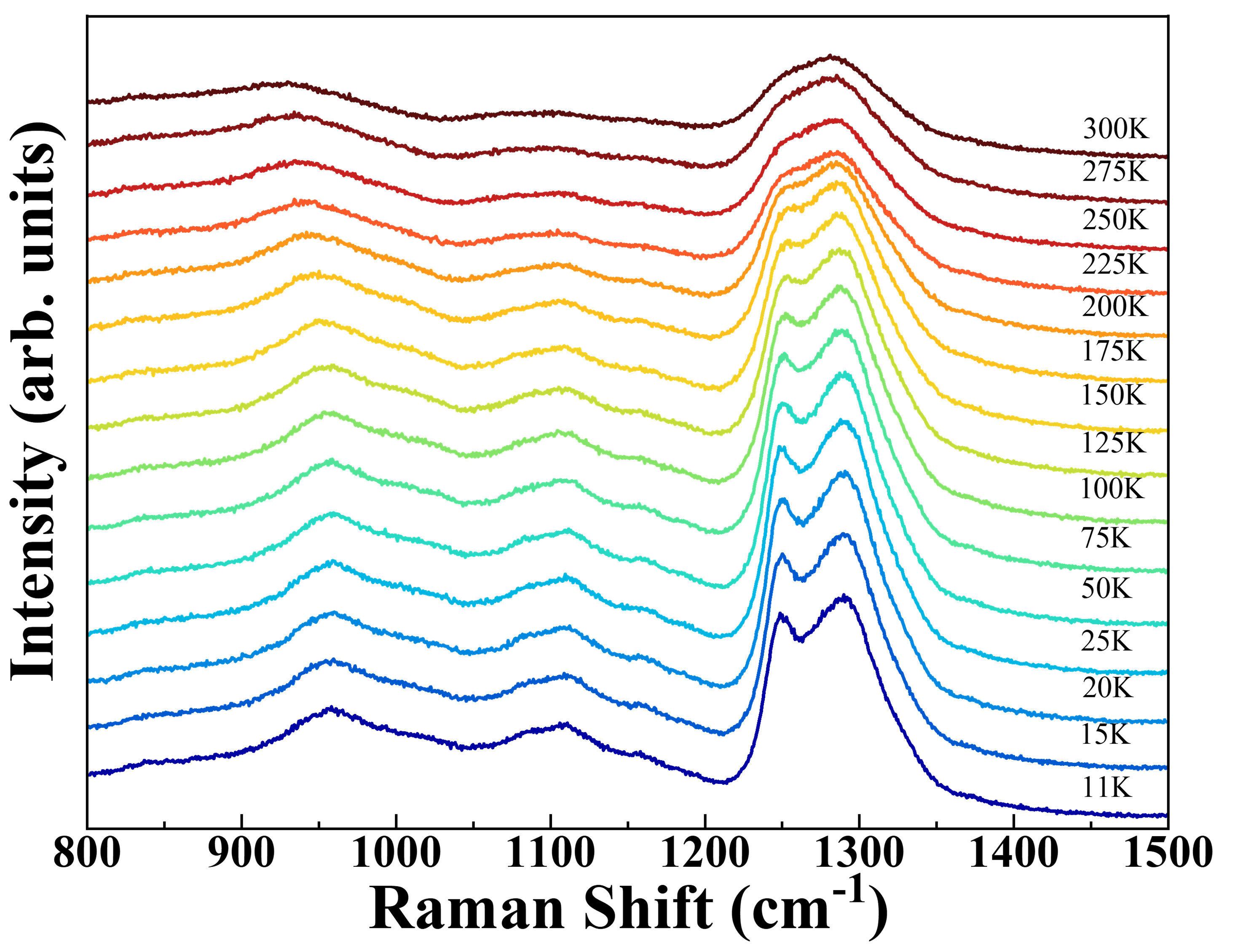}
        \caption{Temperature-dependent Raman scattering spectra of TbFeO$_3$ (TFO) showing the evolution of magnon modes and two-phonon modes in the high-wavenumber region.}
        \label{fig:S6}
    \end{figure}
    
\subsection*{Phenomenological fitting of the high-frequency Raman response}
\begin{figure}[htbp]
    \centering
    \includegraphics[width=\textwidth]{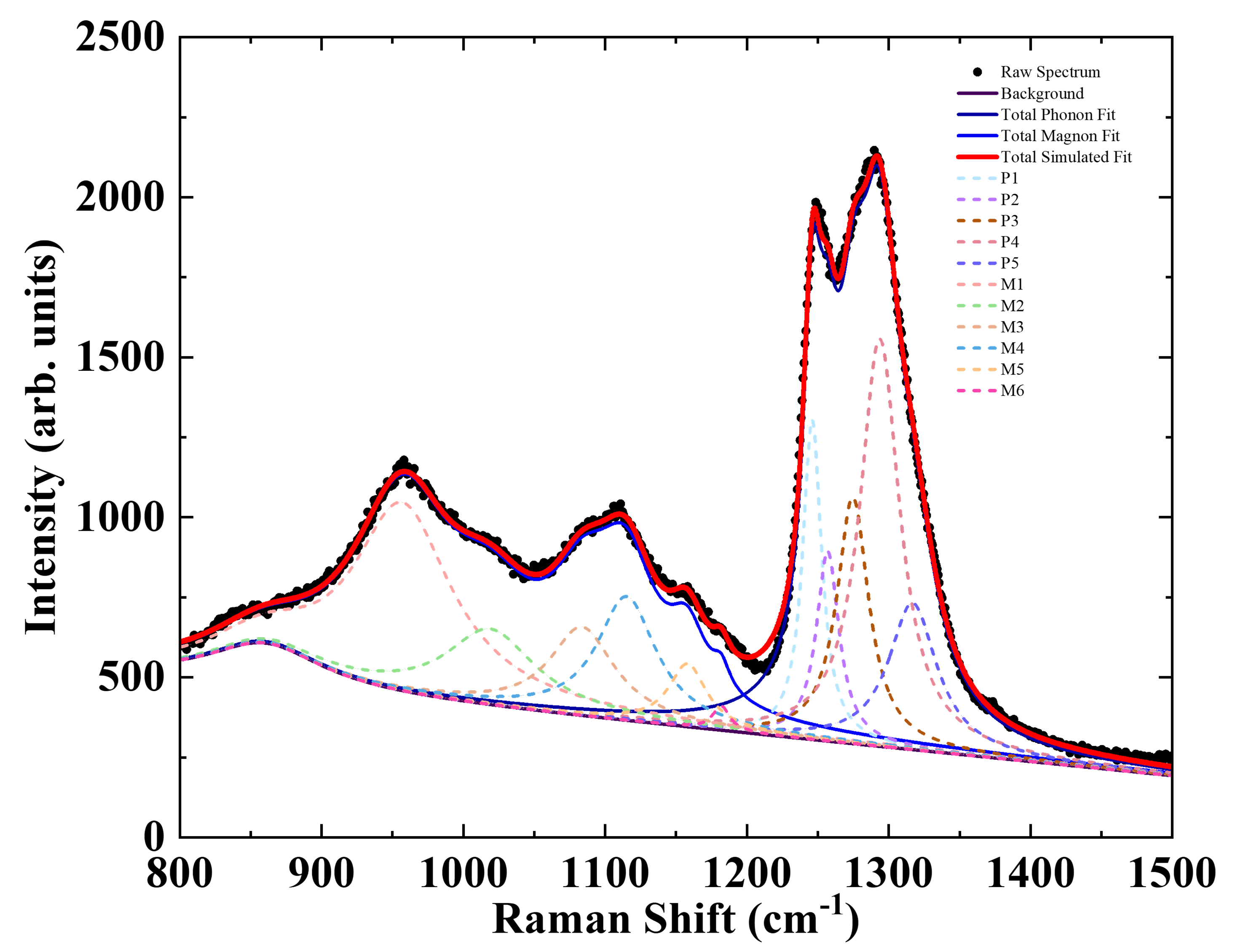}
    \caption{
    Representative phenomenological fit of the high-frequency Raman spectrum of TbFeO$_3$ in the 800--1500~cm$^{-1}$ region at 11~K. The black symbols represent the experimental data, while the solid red curve shows the total fitted profile. The dashed colored curves denote the individual Lorentzian components together with the smooth background. The decomposition is used to separate overlapping multi-particle Raman contributions and is not intended as a one-to-one assignment to individual magnon or phonon branches.
    }
    \label{fig:high_freq_fit}
\end{figure}

\begin{figure}[htbp]
    \centering 
    \includegraphics[width=\textwidth]{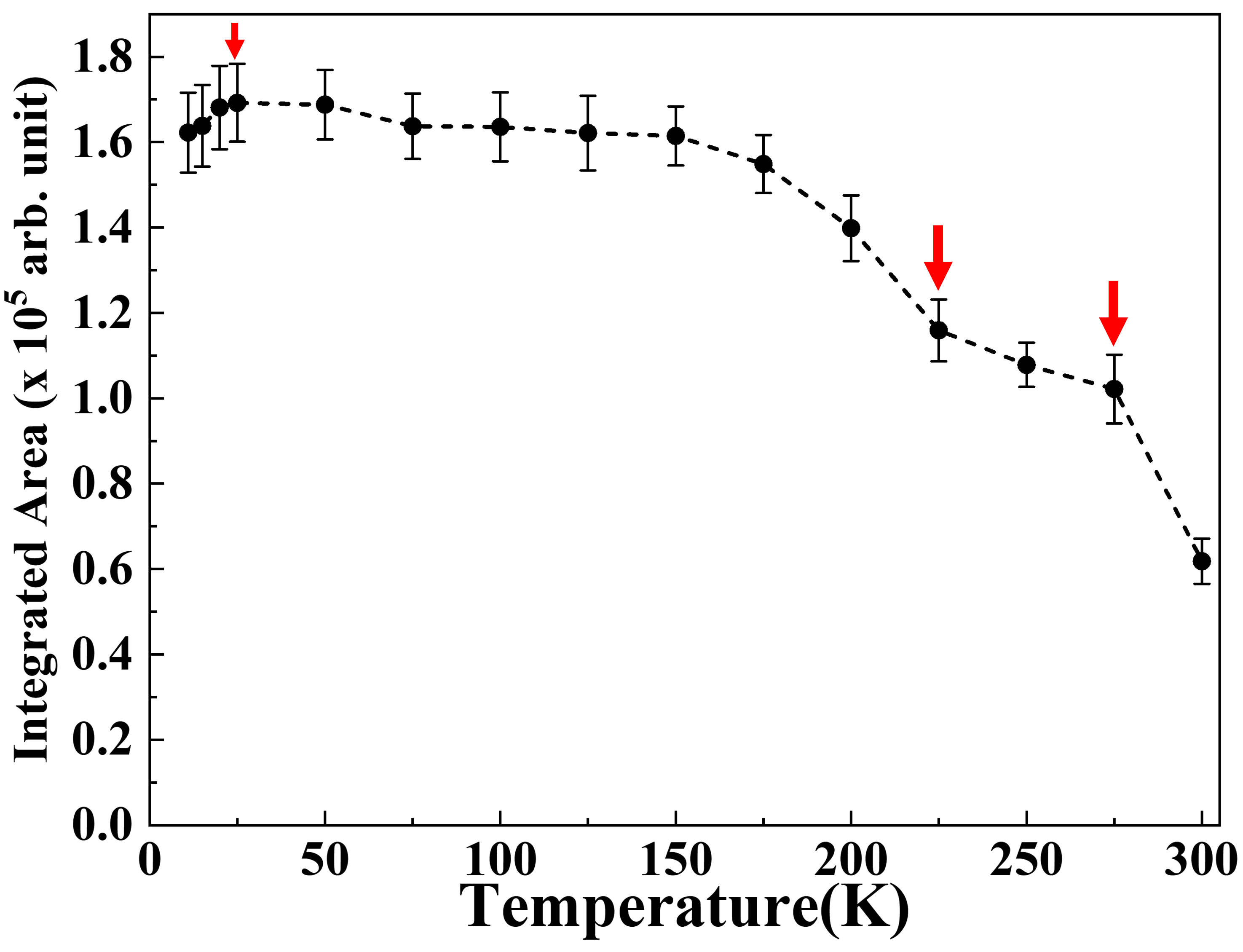}
    \caption{
    Temperature dependence of the integrated spectral weight of the fitted components associated with the  two-magnon-related Raman response in polycrystalline TbFeO$_3$. The second-order phonon contribution was excluded from the integration after the phenomenological fitting.
    }
    \label{fig:areaMagnon}
\end{figure}

To examine the temperature dependence of the weak high-frequency Raman features, the spectra in the 800--1500~cm$^{-1}$ region were fitted phenomenologically using multiple Lorentzian components together with a smooth background. A representative fit of the 11~K spectrum is shown in Fig.~\ref{fig:high_freq_fit}. The black symbols denote the experimental data, the solid red curve represents the total fit, and the dashed curves show the individual fitted components. The broad response in this spectral window contains overlapping contributions from weak magnetic Raman features and a stronger second-order phonon background. Therefore, the Lorentzian components are not assigned uniquely to individual magnon or phonon branches; instead, they provide a practical decomposition of the broad multi-particle Raman response. The same fitting procedure was applied to all measured temperatures. After fitting, the contribution associated with the second-order phonon band was excluded, and the integrated spectral weight of the components assigned to the possible two-magnon-related response was recalculated. The resulting temperature dependence is shown in Fig.~\ref{fig:areaMagnon}. The integrated intensity follows the same overall trend as the raw high-frequency Raman spectral weight discussed in the main text, supporting that the observed temperature evolution is not solely caused by the second-order phonon background. Nevertheless, the comparison with the calculated magnon density of states should be regarded as qualitative, since two-magnon Raman scattering can involve a broad range of magnon-pair states, with enhanced contributions from regions of large joint density of states.

    \begin{figure}[H]
        \centering
        \includegraphics[width=\textwidth]{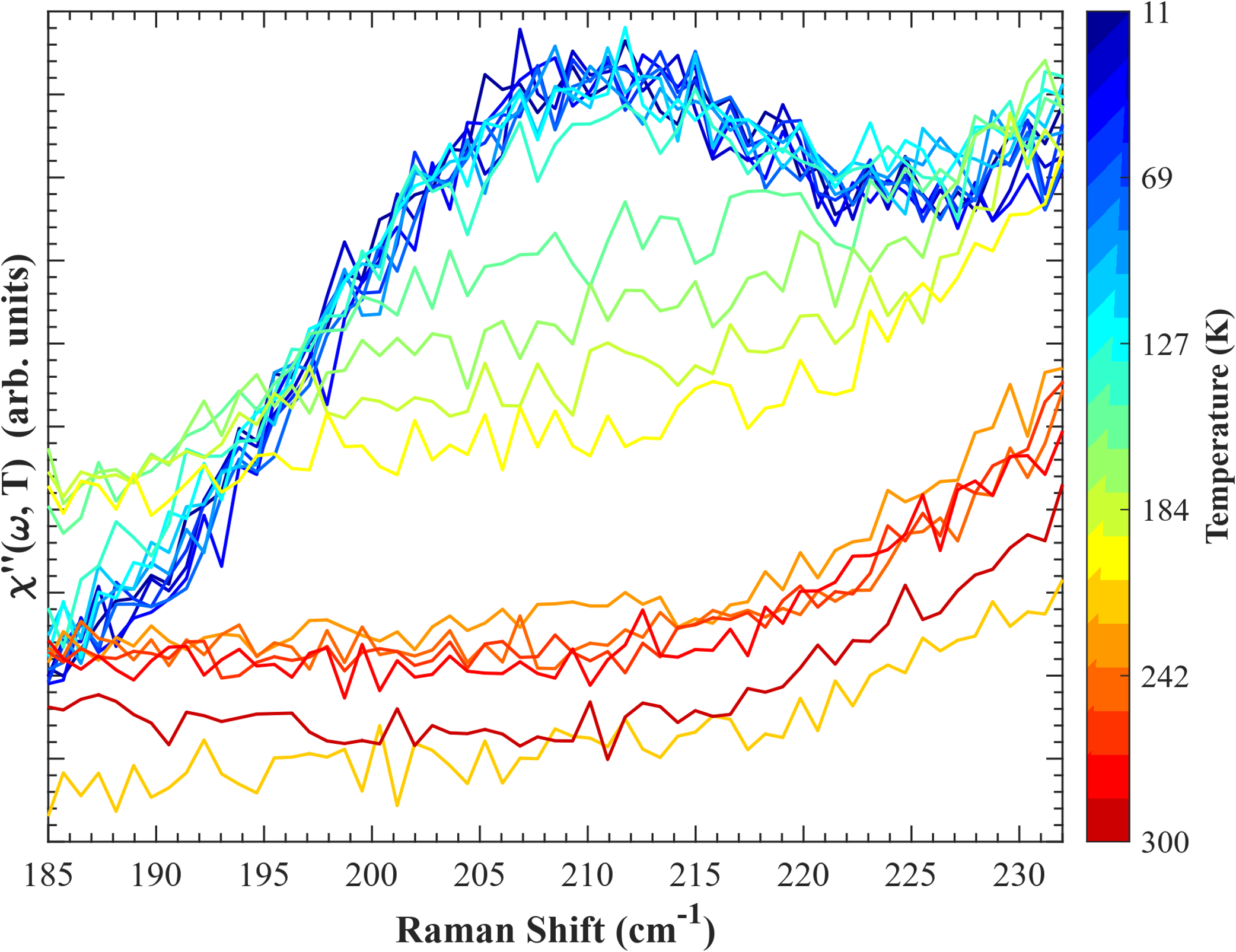}
    \caption{Temperature-dependent Raman scattering spectra of TbFeO$_3$ (TFO) showing the evolution of Raman mode near 206~cm$^{-1}$.}
        \label{fig:S10}
    \end{figure}

\end{document}